\documentstyle[12pt,fleqn]{article}
\topmargin - 0.8cm

\catcode`@=11
\@addtoreset{equation}{section}
\catcode`@=12
\begin{document}
\title{\vskip-1.7cm \bf  Solution of quantum Dirac constraints via path
integral}
\date{}
\author{A.O.Barvinsky}
\maketitle
\hspace{-8mm}{\em
Theory Department, Lebedev Physics Institute and Lebedev Research Center in
Physics, Leninsky Prospect 53,
Moscow 117924, Russia}
%\xpt
\begin{abstract}
%\xpt
The semiclassical solution of quantum Dirac constraints in generic
constrained system is obtained by directly calculating in the one-loop
approximation the gauge field path integral with relativistic
gauge fixing procedure. The gauge independence property of this path
integral is analyzed by the method of Ward identities with a special
emphasis on boundary conditions for gauge fields. The calculations
are based on the known reduction algorithms for functional determinants
extended to gauge theories. The mechanism of transition from relativistic
gauge conditions to unitary gauges, participating in the construction
of this solution, is explicitly revealed. Implications of this result
in problems with spacetime boundaries, quantum gravity and cosmology
are briefly discussed.
\end{abstract}

\section{Introduction}
\hspace{\parindent}
It is well known that the path integral is an efficient tool for
solving the Schrodinger equation. Numerous results in
modern physics in this or that way are related to the path integral
method. The virtue of this method is that it is applicable to any
evolutionary (parabolic) equation and for the Schrodinger equation
gives its unitary evolution operator. At the same time
there exists a class of problems in which the fundamental dynamical
equations are not of a manifestly evolution type. A wide class of such
problems is represented by Dirac quantization of constrained dynamical
systems \cite{Dirac,DW}. In such systems the Schrodinger equation is
supplemented by
quantum Dirac constraints on quantum states, so that the problem amounts
to solving the whole system of equations, not all of
them being of the evolutionary type. Moreover, in the parametrized systems
with a vanishing Hamiltonian there is no independent
Schrodinger equation and their quantum dynamics is encoded in the Dirac
constraints along with their gauge invariance properties. Applications of
the path integral method in this context are much less known and generally
look as follows.

Consider dynamical systems with the canonical action
       \begin{eqnarray}
	S\,[\,q,p,N\,]=\int_{t_-}^{t_+}dt\,
	[\,p_i \dot{q}^i-N^\mu T_\mu(q,p)\,]         \label{00.1}
	\end{eqnarray}
in configuration space of canonical coordinates and momenta $(q,p)=(q^i,p_i)$
and Lagrange multiplyers $N=N^\mu$. The variation of $N^\mu$
leads to nondynamical equations -- the constraints
	\begin{eqnarray}
	T_\mu(q,p)=0.                              \label{00.2}
	\end{eqnarray}
The constraint functions on phase space $T_\mu(q,p)$ belong to the first
class when they satisfy the Poisson bracket algebra
	\begin{eqnarray}
	\{T_\mu,T_\nu\}=U^{\alpha}_{\mu\nu}T_\alpha       \label{00.3}
	\end{eqnarray}
with some structure functions $U^{\alpha}_{\mu\nu}=U^{\alpha}_{\mu\nu}(q,p)$.
This algebra indicates that the theory possesses a local gauge invariance
under the action of canonical transformations of $(q,p)$ generated by
constraints themselves and by certain transformations of Lagrange
multiplyers \cite{FV,BF}.

Dirac quantization of the theory (\ref{00.1}) consists in promoting initial
phase-space variables and constraint functions to the
operator level $(q,p,T_\mu)\rightarrow (\hat{q},\hat{p},\hat{T}_\mu)$ and
selecting the physical states $|\,{\mbox{\boldmath$\Psi$}}\big>$ in the
representation space of $(\hat{q},\hat{p},\hat{T}_\mu)$ by the equation
	$\hat{T}_\mu|\,{\mbox{\boldmath$\Psi$}}\big>=0$
\cite{Dirac,DW,Hen,BarvU,BMarn}. Operators $(\hat{q},\hat{p})$ are subject to canonical
commutation relations $[\hat{q}^k,\hat{p}_l]=i\hbar\delta^k_l$ and the
quantum constraints $\hat{T}_\mu$ as operator functions of
$(\hat{q},\hat{p})$ should satisfy the correspondence principle with
classical $c$-number constraints and be subject to the commutator algebra
	\begin{eqnarray}
	[\hat{T}_\mu,\hat{T}_\nu]=
	i\hbar\hat{U}^{\lambda}_{\mu\nu}\hat{T}_\lambda.  \label{00.5}
	\end{eqnarray}
with certain operator structure functions $\hat{U}^{\lambda}_{\mu\nu}$
standing to the left of operator constraints. This algebra generalizes
(\ref{00.3}) to the quantum level and serves as integrability conditions
for quantum constraints. In the coordinate representation,
	$\big<\,q\,|\,{\mbox{\boldmath$\Psi$}}\big>=
	{\mbox{\boldmath$\Psi$}}(q),\,\,\,\,
	p_k=\hbar\partial/i\partial q^k$,
the latter become the equations on the physical wave function
	\begin{eqnarray}
	\hat{T}_\mu(q,\hbar\partial/i\partial q)\,
	{\mbox{\boldmath$\Psi$}}(q)=0                  \label{00.7}
	\end{eqnarray}
with the differential operators of quantum Dirac constraints
$\hat{T}_\mu(q,\hbar\partial/i\partial q)$.

Note that without loss of generality we did not include in (\ref{00.1})
the Hamiltonian $H(q,p)$ nonvanishing on constraint equations (\ref{00.2}).
By extending the phase space of the theory with extra canonical pair
$(q^0,p_0),\,q^0\equiv t,$ subject to the constraint $p_0+H(q,p)=0$ one can
always reduce the system to the case of action (\ref{00.1}). At the quantum
level this extra Dirac constraint plays the role of the Schrodinger equation
for the wave function of the theory with parametrized time,
${\mbox{\boldmath$\Psi$}}(t,q)={\mbox{\boldmath$\Psi$}}(q^0,q)$,
	\begin{eqnarray}
	\left[\,\frac{\hbar}i\frac{\partial}{\partial q^0}
	+\hat{H}(q,\hbar\partial/i\partial q)\,\right]\,
	{\mbox{\boldmath$\Psi$}}(q^0,q)=0.                \label{00.8}
	\end{eqnarray}
In this sense the dynamical content of any theory can be encoded in the
quantum Dirac constraints of the above type.

The path integral in this context arises as a special solution of quantum
constraints (\ref{00.7}) in the form of the two-point kernel
${\mbox{\boldmath$K$}}(q,q')$ -- the analogue of the two-point evolution
operator for the Schrodinger equation \cite{Leutw,BAO}
	\begin{eqnarray}
	\hat{T}_{\mu}(q,\hbar\partial/i\partial q)\,
	{\mbox{\boldmath$K$}}(q,q')=0.                     \label{0.0}
	\end{eqnarray}
Similarly to the theory of the Schrodinger equation this is a path integral
over configuration space variables $(q(t),N(t))$ in
(space)time domain $t_-<t<t_+$ with the boundary conditions related to
the arguments of this kernel $q(t_+)=q$, $q(t_-)=q'$. However, in view of
the non-evolutionary nature of equations (\ref{0.0}) the path integral
construction here is much less straightforward than in the Schrodinger case.

Mainly the motivation for such applications comes from quantum gravity theory
\cite{Leutw,BAO,HH,barvin,HalHar}. In this theory (\ref{00.1}) is a canonical
form of the Einstein action with $q^i=g_{ab}({\bf x})$ -- 3-metric
coefficients and $N^\mu\sim ((-g_{00})^{-1/2}({\bf x}), g_{0a}({\bf x}))$
-- lapse and shift functions in (3+1)-foliation of spacetime \cite{ADM},
(\ref{00.2}) are the gravitational Hamiltonian and momentum constraints
and (\ref{00.7}) represents the system of Wheeler-DeWitt equations on the
cosmological wavefunction
${\mbox{\boldmath$\Psi$}}(q)={\mbox{\boldmath$\Psi$}}\,[\,g_{ab}({\bf x})\,]$
\cite{DW}. The construction of the path integral here is persuing two
main goals. The first goal is a two-point solution for the Wheeler-DeWitt
equations (\ref{0.0}) \cite{Leutw,BAO,barvin,HalHar}, which in view of
parametrized nature of time encodes the dynamical information. The second goal
consists in specifying the distinguished
cosmological quantum state -- the model for initial conditions in quantum
cosmology of the early universe \cite{HH}. The first formulation of the path
integral for the two-point solution of Wheeler-DeWitt equations belongs
to H.Leutwyler \cite{Leutw}. It was only qualitatively correct because
at that time the
structure of gauge fixing procedure and the role of ghost fields in the path
integral have not yet been understood. The path integral with unitary
gauge fixing procedure and exhaustive set of boundary conditions was
later proposed in \cite{BAO}. Then this canonical path integral
was converted to the spacetime covariant form of the functional integral
over Lagrangian variables in relativistic gauges \cite{barvin,HalHar}.

Of course, these results encorporate a well-known statement on
equivalence of the canonical and covariant quantizations pioneered in
\cite{FV}. In contrast to this, the works \cite{BAO,HH,barvin,HalHar}
were focused on the nontrivial boundary
conditions in spacetime. Correct treatment of these boundary conditions
leads to the proof that this path integral solves the quantum Dirac
constraints (\ref{0.0}). However, this proof given in
\cite{BAO,HH,barvin0,HalHar} has a formal nonperturbative nature and does
not even allow one to fix the operators of quantum constraints.
One can only infer from this proof that these operators satisfy the
correspondence principle with their classical counterparts and have quantum
corrections which in a rather uncontrollable way depend on the calculational
method for a path integral \cite{barvin0}. Thus, no check of the solution
to quantum constraints was thus far given by direct calculations of the
path integral. The goal of this paper is to perform such a check.
This will be done for generic systems subject to first class constraints
of the above type in the one-loop approximation of semiclassical expansion.

The nature of this check is certainly a comparison of two results: one
obtained by solving the equations (\ref{0.0}) and another by calculating
the path integral. In the one-loop approximation for the Schrodinger
equation such a comparison is based on the Pauli-Van Vleck-Morette
formula \cite{Morette} and the reduction method for functional determinants
\cite{reduct}. The
preexponential factor in the subleading semiclassical order
is given by the Van Vleck determinant -- the solution of the
corresponding continuity equation. On the other hand, it is given by the
functional determinant of the inverse propagator of the theory -- the
contribution resulting from the one-loop (gaussian) approximation for
the path integral. The equality of these two expressions follows from the
reduction method for functional determinants considered in \cite{reduct}.
The semiclassical solution of quantum Dirac constraints in terms of
the modified Van Vleck determinant was built in the author's paper
\cite{GenSem,BKr}. Here we reproduce this solution from the
Faddeev-Popov path integral of ref.\cite{barvin}.

The paper is organized as follows. In Sect.2 we recapitulate the
semiclassical solution of quantum Dirac constraints of refs. \cite{GenSem,BKr}
with a particular emphasis on unitary gauges participating in its
construction. Sect.3 serves as a link between the canonical formalism
of the constrained system and its Lagrangian version with local gauge
invariances presented in (space)time covariant form. The corresponding
Faddeev-Popov path integral in relativistic gauges constructed as a
solution of quantum Dirac constraints in \cite{barvin} is presented
in Sect.4 along with its semiclassical expansion up to the one-loop order.
Sect.5 deals with the mechanism of its gauge independence based on Ward
identities and, in particular, gauge independence of its local measure
cancelling the strongest power divergences. In Sect.6 we calculate the
contribution of the gauge field functional determinant by deriving the
algorithm of its reduction to the Van Vleck elements of the semiclassical
solution of Sect.2. Similar calculations are performed for the
ghost field determinant in Sect.7. They accomplish the proof of the main
result -- path integral derivation of the solution to quantum Dirac
constraints. In concluding section we discuss implications of this
result regarding the aspects of gauge invariance in quantum cosmology
and Euclidean quantum gravity.

The final remark, which is in order here, concerns our notations. Throughout
the paper we use condensed DeWitt notations \cite{DW} which allow one in a
manageable form to handle both the quantum mechanical and field theoretical
problems within the language of the above type. We imply that the
range of indices of $q^i$ and $N^\mu$
	\begin{eqnarray}
	i=1,...,n,\,\,\,\mu=1,...,m,       \label{00.9}
	\end{eqnarray}
in field models formally extends to infinite dimensionalities of phase
space $n$ and space of gauge transformations $m$, and these {\it canonical}
condensed indices together with discrete tensor labels carry also
continuous labels of spatial coordinates ${\bf x}$ (or certain discrete
numbers when the fields are expanded in the basis of some
countable set of spatial harmonics). Contraction of these indices will
include spatial integration (or the corresponding infinite
summation)\footnote
{
This formal approach does not certainly incorporate a rigorous handling
of ultraviolet infinities and possible quantum anomalies which go beyond
the scope of this paper. The justification of this approach is based on
the fact that our considerations serve as a bridge between
the manifestly noncovariant Dirac quantization and its Lagrangian
path integral counterpart that can be cast into manifestly covariant form.
It is the latter formalism that must be covariantly regulated to give
a physically reasonable framework for renormalization and quantum
anomalies.
}.
Starting from Sect. 3 we shall also need the {\it covariant} condensed
indices including the time label and allowing one to represent the covariant
(space)time operations of integration and differentiation in the form of
a simple contraction.

\section{Canonical solution of quantum constraints: unitary gauges}
\hspace{\parindent}
The solution of quantum constraints (\ref{0.0}) found in
\cite{GenSem,BKr,BarvU}
	\begin{eqnarray}
	{\mbox{\boldmath$K$}}(q,q')=
	{\mbox{\boldmath$P$}}(q,q')\,
	e^{\,{}^{\textstyle{\frac i\hbar
	{\mbox{\boldmath$S$}}(q,q')}}}            \label{0.1}
	\end{eqnarray}
contains the Hamilton-Jacobi function ${\mbox{\boldmath$S$}}(q,q')$
which determines the tree-level approximation and the preexponential
factor ${\mbox{\boldmath$P$}}(q,q')$ accumulating loop corrections. The
system of Hamilton-Jacobi equations for ${\mbox{\boldmath$S$}}(q,q')$
	\begin{eqnarray}
	T_\mu(q,\partial {\mbox{\boldmath$S$}}/\partial q)=0  \label{0.2}
	\end{eqnarray}
enforces the quantum constraints in the leading order of $\hbar$-expansion.
For their operator realization proposed in \cite{GenSem,BKr,BarvU} the
subleading (one-loop) order yields the following quasi-continuity equations
for the preexponential factor ${\mbox{\boldmath$P$}}(q,q')$ (see also
\cite{Kiefer1} in the gravitational context)
	\begin{eqnarray}
	&&\frac\partial{\partial q^i}
	(\nabla^i_\mu {\mbox{\boldmath$P$}}^2)=
	U^{\lambda}_{\mu\lambda} {\mbox{\boldmath$P$}}^2,  \label{0.3}\\
	&&\nabla^i_\mu\equiv\left.
	\frac{\partial T_\mu}{\partial p_i}
	\right|_{\;\textstyle p=\partial
	{\mbox{\boldmath$S$}}/\partial q}.                    \label{0.4}
	\end{eqnarray}

A particular solution of eqs.(\ref{0.2}) - (\ref{0.3}) corresponds to the
choice of the principal Hamilton function for ${\mbox{\boldmath$S$}}(q,q')$
-- the classical action (\ref{00.1}) calculated at the extremal of equations
of motion $g=g(\,t\,|q,q')$ joining the configuration space points
$q$ and $q'$. Together with eq.(\ref{0.2}) this function also satisfies the
Hamilton-Jacobi equations with respect to its second argument
	\begin{eqnarray}
	T_\mu(q',-\partial {\mbox{\boldmath$S$}}/\partial q')=0. \label{0.5}
	\end{eqnarray}

Similarly to the WKB theory of non-gauge quantum systems, this function
gives rise to a special one-loop preexponential factor \cite{GenSem,BKr,BarvU}
which is a generalization of the Pauli-Van Vleck-Morette
formula \cite{Morette} -- calculating the determinant of the matrix of
second order derivatives of the principal Hamilton function with respect to
its two arguments. However, in contrast with non-gauge theories this
factor cannot be directly constructed in terms of this determinant, because
in view of eqs.(\ref{0.2}) and (\ref{0.5}) the matrix
	\begin{eqnarray}
	{\mbox{\boldmath$S$}}_{ik'}=
	\frac{\partial^2{\mbox{\boldmath$S$}}(q,q')}
	{\partial q^i\, \partial q^{k'}},             \label{0.6}
	\end{eqnarray}
has left and right zero-eigenvalue eigenvectors \cite{GenSem,BKr,BarvU}
and, therefore, is degenerate
	\begin{eqnarray}
	&&\nabla^i_\mu{\mbox{\boldmath$S$}}_{ik'}=0,    \label{0.7}\\
	&&{\mbox{\boldmath$S$}}_{ik'}\nabla^{k'}_\nu=0,\,\,\,\,
	\nabla^{k'}_\nu\equiv\left.
	\frac{\partial T_\nu}{\partial p_{k'}}
	\right|_{\;\textstyle p=-\partial
	{\mbox{\boldmath$S$}}/\partial q'}.             \label{0.8}
	\end{eqnarray}

The construction of the one-loop preexponential factor in terms of
this degenerate matrix is equivalent to the Faddeev-Popov gauge-fixing
procedure for gauge theories. It consists in adding the
``gauge-breaking'' term bilinear in ``gauge conditions''
$X^\mu_i$ and $X^\nu_{k'}$ -- two sets of arbitrary covectors at the
configuration space points $q$ and $q'$
	\begin{eqnarray}
	D_{ik'}={\mbox{\boldmath$S$}}_{ik'}+
	X^\mu_i C_{\mu\nu} X^\nu_{k'}.             \label{0.9}
	\end{eqnarray}
This allows one to replace the degenerate matrix
${\mbox{\boldmath$S$}}_{ik'}$ by the new invertible matrix $D_{ik'}$,
provided that the gauge-fixing matrix $C_{\mu\nu}$ is also invertible
and these gauge covectors produce invertible ``Faddeev-Popov operators''
\cite{Fad}
	\begin{eqnarray}
	&&J^\mu_\nu=X^\mu_i\nabla^i_\nu,\,\,\,
	J\equiv{\rm det}J^\mu_\nu\neq 0,             \label{0.10} \\
	&&J'^{\mu}_{\,\nu}=X^\mu_{i'}\nabla^{i'}_\nu,\,\,\,
	J'\equiv{\rm det}J'^{\mu}_{\,\nu}\neq 0.             \label{0.11}
	\end{eqnarray}

In terms of these objects the solution of the continuity equations (\ref{0.3})
is given by the following expression \cite{GenSem,BKr,BarvU}
	\begin{eqnarray}
	{\mbox{\boldmath$P$}}=
	\left[\frac{{\rm det}\,D_{ik'}}
	{JJ'\,{\rm det}\,C_{\mu\nu}}\right]^{1/2},      \label{0.12}
	\end{eqnarray}
which can be regarded as an analogue of the one-loop expression for
the effective action of gauge field theory -- the contribution of
gauge fields ${\rm det}\,D_{ik'}$ partly compensated by the contribution of
ghosts $J$ and $J'$. This compensation makes the prefactor independent of
the introduced arbitrary elements of gauge-fixing procedure
$(X^\mu_i,X^\nu_{k'}, C_{\mu\nu})$ -- the analogue of on-shell gauge
independence in gauge field theory. The mechanism of this gauge independence
is based on the ``Ward identities'' for the gauge field ``propagator''
(\ref{0.9})
	\begin{eqnarray}
	C_{\mu\nu}\,X^\nu_{k'}\,D^{-1\;k'i}=
	J^{-1\,\nu}_\mu \nabla^i_\nu,                      \label{0.13}
	\end{eqnarray}
easily obtained by contracting (\ref{0.9}) with $\nabla^i_\mu$ and using
(\ref{0.7}). The use of these identities shows that arbitrary
variations of the quantities $(X^\mu_i,X^\nu_{k'},C_{\mu\nu})$ in
the one-loop prefactor vanish due to the cancellation of terms coming from
the gauge field ${\rm det}\,D_{ik'}$ and ghost $JJ'$ determinants.

The nature of covectors $(X^\mu_i,X^\nu_{k'})$ as matrices of gauge conditions
does not only follow from the fact that they remove the degeneracy of the
matrix ${\mbox{\boldmath$S$}}_{ik'}$ caused by gauge invariance of the theory.
As shown in \cite{GenSem,BKr} the quantum Hamiltonian reduction of the
kernel ${\mbox{\boldmath$K$}}(q,q')$ with the one-loop factor
(\ref{0.12}) leads to the unitary evolution operator in the physical sector
defined by the {\it unitary} gauge conditions
	\begin{eqnarray}
	X^\mu(q,t)=0,                                \label{0.14}
	\end{eqnarray}
such that
	\begin{eqnarray}
	X^\mu_i=\frac{\partial X^\mu}{\partial q^i}.     \label{0.15}
	\end{eqnarray}
The unitary gauge conditions are imposed only on phase space variables of
the theory $(q,p)$ (in this case only on coordinates $q$)\footnote
{
Explicit time dependence of unitary gauge conditions is necessary in the
theories with parametrized time in order to have nontrivial time evolution
with a nonvanishing physical Hamiltonian \cite{BarvU,BKr}.
}.
In contrast with {\it relativistic} gauges involving Lagrange
multiplyers, they manifestly incorporate unitarity and do not give rise to
propagating ghosts. The price one usually pays for manifest unitarity is
the absence of manifest covariance, that can be restored by going over to
relativistic gauges. Thus, the solution (\ref{0.12}) is obtained by directly
solving the quantum Dirac constraints within the framework of unitary
gauge conditions. In what follows we show that the same solution can be
obtained by a direct calculation of the path integral in the relativistic
gauge, and this derivation as a byproduct will establish the relation
between unitary and relativistic gauge conditions.

\section{Lagrangian versus canonical formalisms}
\hspace{\parindent}
As a first step towards the covariant path integral in the relativistic gauge
let us consider the Lagrangian formalism of the theory with the canonical
action (\ref{00.1}). For this we introduce the collective notation for the full
set of {\em Lagrangian} configuration space variables -- canonical coordinates
and Lagrange multiplyers\footnote
{
In the context of Einstein gravity theory this collection of ten fields
$g^a\sim g_{\alpha\beta}({\bf x},t)$ comprises the whole set of spacetime
metric coefficients taken in a special parametrization adjusted to
(3+1)-splitting: $q^i=g_{ab}({\bf x},t),\,
N^\alpha\sim g_{0\alpha}({\bf x},t)$.
}
	\begin{eqnarray}
	&&g^a=(q^i(t),N^\mu(t)).     \label{1.0}
	\end{eqnarray}
In what follows we shall need also (space)time condensed DeWitt notations
in which the index $a$ includes not
only the spin labels and spatial coordinates ${\bf x}$ but also the time
variable $t$, and the contraction of these indices implies the time
integration (as mentioned in Introduction we shall call them
{\it covariant}). In these notations the action has the form
	\begin{eqnarray}
	S\,[\,g\,]=\int_{t_-}^{t_+}dt\,L(q,\dot{q},N) \label{1.2}
	\end{eqnarray}
with the Lagrangian which does not involve time derivatives of the
Lagrange multiplyers $N=N^\mu(t)$. This Lagrangian is related to
the integrand of the canonical action (\ref{00.1})
	\begin{eqnarray}
	L(q,\dot{q},N)=\left.\left( p_i\dot{q}^i-
	N^{\mu}T_\mu (q,p)\right)
	\right|_{\,p=p^0(q,\dot{q},N)}                   \label{1.3}
	\end{eqnarray}
by the substitution of the expression for the canonical momentum
$p_i^0(q,\dot{q},N)$ in terms of the velocities $\dot{q}$
	\begin{eqnarray}
	p_i^0(q,\dot{q},N)=\frac{\partial L(q,\dot{q},N)}
	{\partial\dot{q}^i}                                 \label{1.4}
	\end{eqnarray}
which is a solution of the canonical equation of motion
	\begin{eqnarray}
	\dot{q}{}^i=N^\mu\frac{\partial T_\mu(q,p)}
	{\partial p_i}.                                     \label{1.5}
	\end{eqnarray}

The classical action is invariant under gauge transformations with local
(arbitrary time and space dependent) parameters $f^\mu=f^\mu(t)$
vanishing on spacetime boundary $f^\mu(t_\pm)=0$. For the
canonical action (\ref{00.1}) these transformations are canonical
and, therefore, ultralocal in time for phase space variables, but involve
the time derivative of the gauge parameter for Lagrange multiplyers \cite{FV}
	\begin{eqnarray}
	&&\delta q^i=\{q^i,T_\mu\} f^\mu,\,\,\,
	\delta p_i=\{p_i,T_\mu\} f^\mu,                \label{1.6}\\
	&&\delta N^\mu=\dot{f}^\mu
	-U^{\mu}_{\alpha\nu}N^\alpha f^\nu.             \label{1.7}
	\end{eqnarray}
Thus the first class constraints $T_\mu(q,p)$ serve as generators of
infinitesimal gauge transformations on the phase space of $(q,p)$.
Note, in particular, that the transformations of phase space
{\em coordinates} $q$ when restricted to the Lagrangian surface of
the Hamilton-Jacobi function, $p=\partial{\mbox{\boldmath$S$}}/\partial q$,
read
	\begin{eqnarray}
	\delta q^i=\nabla^i_\mu f^\mu             \label{1.8}
	\end{eqnarray}
in terms of the vector field (\ref{0.4}).

These gauge invariance transformations in the Lagrangian formalism
take the form of the infinitesimal transformations of configuration space
variables
	\begin{eqnarray}
	&&\delta g^a=R^a_\mu f^\mu,              \label{1.9}\\
	&&R^a_\mu\,\frac{\delta S\,[\,g\,]}
	{\delta g^a}=0.                          \label{1.10}
	\end{eqnarray}
with the {\it Lagrangian} generators $R^a_\mu=(R^i_\mu,R^\alpha_\mu)$. Here
we use covariant condensed notations in which the index
$a$ includes time, and its contraction implies the time integration.
Thus, $R^a_\mu$ forms a delta-function type kernel with two entries
$a\rightarrow (a,t),\,\mu\rightarrow (\mu,t')$
	\begin{eqnarray}
	R^a_\mu=R^a_{\;\mu}(d/dt)\,\delta(t-t'),   \label{1.11}
	\end{eqnarray}
where $R^a_{\;\mu}(d/dt)$ denotes the differential (or ultralocal
multiplication) operator acting on the first argument of the delta function.
In view of the relation (\ref{1.3}) between the canonical and Lagrangian
formalisms different components of this kernel follow from the
invariance transformations (\ref{1.6})-(\ref{1.7}) \cite{FV}\footnote
{
The equivalence of the full set of transformations (\ref{1.6})-(\ref{1.7})
(including transformations of momenta) to (\ref{1.9})-(\ref{1.13}) holds
only up to terms proportional to equations of motion \cite{FV}. Only up to
such terms holds the equality $\delta p\,|_{\,p^0(q,\dot{q},N)}=
\delta p^0(q,\dot{q},N)$. For us, however, it is only important to know
that Lagrangian gauge transformations can be obtained from the subset of
transformations of coordinates $q^i$ and Lagrange multiplyers $N^\mu$ in
canonical theory.
}
	\begin{eqnarray}
	&&R^i_\mu=\delta(t-t')\left.
	\frac{\partial T_\mu}{\partial p_i}
	\,\right|_{p=p^0(q,\dot{q},N)},           \label{1.12}\\
	&&R^\alpha_\mu=\left(\delta^\alpha_\mu\frac d{dt}
	-U^\alpha_{\lambda\mu}
	N^\lambda\right)\delta(t-t').             \label{1.13}
	\end{eqnarray}
The distinguished role of the Lagrange multiplyers here manifests itself
in the fact that only the $a=\alpha$ component of (\ref{1.11}) forms the
first order differential operator while the other components are ultralocal
in time. In particular, the transformation (\ref{1.12}) is a Lagrangian
form of (\ref{1.8}).

In what follows we shall have to use condensed notations of both
canonical and covariant nature. For brevity we shall not introduce
special labels to distinguish between them. As a rule,
when the time argument is explicitly written we shall imply that the
corresponding condensed index or indices are canonical, i.e. they
contain only spin labels and spatial coordinates and their contraction
does not involve implicit time integrals. For example, the left-hand
side of gauge identities (\ref{1.10}) can be written down in the form
	\begin{eqnarray}
	R^a_\mu\,\frac{\delta S\,[\,g\,]}
	{\delta g^a}=R^{\,\;a}_\mu(d/dt)\,
	\frac{\delta S\,[\,g\,]}{\delta g^a(t)},
	\,\,\,\mu\rightarrow (\mu,t),                  \label{1.14}
	\end{eqnarray}
where the time integral implicit in the contraction of the {\it covariant
condensed} index $a$ removed the delta function contained in $R^a_\mu$
and, thus, the result boiled down to the action of the differential
operator $R^{\,\;a}_\mu(d/dt)$ on $\delta S\,[\,g\,]/\delta g^a(t)$. This
operator obviously differs from that of eq.(\ref{1.11}) by the functional
transpositon -- integration by parts, because in contrast with (\ref{1.11})
it acts on test function with respect to upper condensed index $a$. This
fact is indicated by the order of operator indices reversed relative
to eq. (\ref{1.11}).

Another important distinction between these two types of condensed
notations concerns functional derivatives. We shall always reserve functional
variational notations $\delta/\delta g^a\equiv\delta/\delta g^a(t)$ for
variational derivatives with respect to functions of time, while the
variational derivatives with respect to functions of spatial coordinates
will be denoted by partial derivatives. For example, in the gravitational
context we have
$\delta/\delta g^a\equiv\delta/\delta g_{\alpha\beta}({\bf x},t)$
vs $\partial/\partial q^i\equiv\delta/\delta g_{ab}({\bf x})$.

\section{The path integral in relativistic gauge}
\hspace{\parindent}
The path integral solution of quantum Dirac constraints was constructed in
\cite{barvin}. This is a functional integral over the gauge fields
$g^a$ and grassman ghost fields $C=C^\mu$ and $\bar{C}=\bar{C}_\nu$
with a conventional Faddeev-Popov gauge fixing procedure
	\begin{equation}
	{\mbox{\boldmath$K$}}(q_+,q_-)=
	\int Dg\,\mu[\,g\,]\,DC\,D\bar{C}
	\exp\frac i\hbar \left\{\left(S\,[\,g\,]-
	\frac 12\chi^\mu c_{\mu\nu}\chi^\nu\right)
	+\bar{C}_\mu Q^\mu_\nu C^\nu\right\}.         \label{2.1}
	\end{equation}
The total action in the exponential here contains the gauge fixed
classical and ghost actions. The gauge fixed classical action
	\begin{eqnarray}
	 S_{\rm gf}[\,g\,]=S\,[\,g\,]-
	\frac 12\,\chi^\mu c_{\mu\nu}\chi^\nu,     \label{2.2}
	 \end{eqnarray}
includes the gauge breaking term
	\begin{eqnarray}
	\frac 12\,\chi^\mu c_{\mu\nu}\chi^\nu=\frac 12\,
	\int_{t_{-}}^{t_{+}}dt\,\chi^\mu(g,\dot g)\,
	c_{\mu\nu}\chi^\nu(g,\dot g)                   \label{2.3}
	\end{eqnarray}
quadratic in the {\it relativistic} gauge conditions
	\begin{eqnarray}
	&&\chi^\mu=\chi^\mu(g,\dot g),           \label{2.4}\\
	&&a^\mu_\nu=-\frac{\partial\chi^\mu}
	{\partial{\dot N}^\nu},\,\,\,
	{\rm det}\,a^\mu_\nu \neq 0.             \label{2.5}
	\end{eqnarray}
An important distinction of these gauge conditions is that they involve
the velocities of {\it all} Lagrange multiplyers which results in
propagating nature of all gauge and ghost fields. The corresponding
Faddeev-Popov operator $Q^\mu_\nu$, the kernel of the ghost action
bilinear in ghost fields $C$ and $\bar{C}$
	\begin{eqnarray}
	Q^\mu_\nu=\frac{\delta\chi^\mu}{\delta g^a}R^a_\nu, \label{2.6}
	\end{eqnarray}
turns out to be a second order differential operator (cf. eqs.(\ref{1.13})
and (\ref{2.5}))
	\begin{eqnarray}
	Q^\mu_{\;\,\nu}(d/dt)\delta(t-t')=
	\left(-a^\mu_\nu\,
	{d^2}/{dt^2}+...\right)\delta(t-t').          \label{2.7}
	\end{eqnarray}

An important ingredient of the path integral (\ref{2.1}) is the local
integration measure $\mu[\,g\,]$ which is built of the Hessian matrix
of the classical Lagrangian $G_{ik}$ and invertible matrix $c_{\mu\nu}$
fixing the gauge in (\ref{2.3})
 	\begin{eqnarray}
	&&\mu[\,g\,]=\prod_{t}[\,{\rm det}\,G_{ik}(t)\;
	{\rm det}\,c_{\mu\nu}(t)\,]^{1/2}
	\equiv\,\left(\,{\rm Det}\,G_{ik}\;
	{\rm Det}\,c_{\mu\nu}\right)^{1/2},       \label{2.8}\\
	&&G_{ik}=\frac{\partial^2 L}
	{\partial\dot q^{i}\partial\dot q^{k}}.   \label{2.9}
	\end{eqnarray}
Note that the symbol $\rm det$ in (\ref{2.5}) and in the middle part of
eq.(\ref{2.8}) denotes the determinants of matrices acting in the space
of canonical condensed indices. The products over time points
of local factors ${\rm det}\,G_{ik}(t)$ and ${\rm det}\,c_{\mu\nu}(t)$
can be regarded as determinants of higher functional dimensionality
if we redefine these matrices as time-ultralocal operators acting in
the space of condensed covariant indices $G_{ik}\equiv G_{ik}\delta(t-t')$
and $c_{\mu\nu}\equiv c_{\mu\nu}\delta(t-t')$. We shall denote such
functional determinants for both ultralocal and differential operators
in time by ${\rm Det}$. Thus the right-hand side of eq.(\ref{2.7}) is
formally written down in terms of such (purely divergent) functional
determinant of the ultralocal operator. For the Hessian matrix it looks
like
	\begin{eqnarray}
	{\rm Det}\,G_{ik}\delta(t-t')=\exp\,
	\left\{\,\delta(0)\int_{t_-}^{t_+}dt\,
	{\rm ln}\,{\rm det}\,G_{ik}\,\right\}.      \label{2.10}
	\end{eqnarray}
Similar definition holds for the ultralocal gauge-fixing matrix\footnote
{
We don't consider here the case when $c_{\mu\nu}$ is a differential or
nonlocal operator. The only difference in this case is that it is no longer a
part of the local measure. It contributes to the path integral a
nontrivial functional determinant usually represented as a gaussian integral
over auxiliary Nielsen-Kallosh ghosts.
}.

The final igredient which accomplishes the definition of the path integral
(\ref{2.1}) is a specification of boundary conditions on integration
variables. Integration in (\ref{2.1}) runs over field histories with
fixed values of canonical {\it coordinates} and ghost fields at $t_{\pm}$
	\begin{eqnarray}
	q^i(t_\pm)=q^i_\pm,\,\,\,C^\mu(t_\pm)=0,
	\,\,\,\bar{C}_\nu(t_\pm)=0,                  \label{2.11}
	\end{eqnarray}
the coordinates $q^i_\pm$ being the arguments of the two-point kernel
${\mbox{\boldmath$K$}}(q_+,q_-)$. On the contrary, the boundary values of
Lagrange multiplyers are integrated over in the infinite range
	\begin{eqnarray}
	-\infty<N^\mu(t_\pm)<+\infty.              \label{2.12}
	\end{eqnarray}

By using these boundary conditions, invariant with respect to BRST
transformations of the total action in (\ref{2.1}), one can show the
gauge independence of the path integral and also give a formal proof
that this integral solves quantum Dirac constraints in the coordinate
representation of canonical commutation relations for $q=q_+$ and
$p_+=\hbar\partial/i\partial q_+$. This proof is based on an obvious
consequence of the integration range for $N^\mu(t_+)$
	\begin{eqnarray}
	\int Dg\,DC\,D\bar{C}\,
	\frac\delta{\delta N^\mu(t_+)}\,\Big(\,...\,\Big)=0,       \label{2.13}
	\end{eqnarray}
where ellipses denote the full integrand of the path integral (\ref{2.1}).
The functional differentiation here boils down in the main to the
deexponentiation of the constraint
	\begin{eqnarray}
	\frac{\delta S\,[\,g\,]}{\delta N^\mu(t_+)}=
	-T_\mu(q(t_+),p(t_+))\,
	\Big|_{\,p=p^0(q,\dot{q},N)}         \label{2.14}
	\end{eqnarray}
(the differentiation of gauge-breaking and ghost terms gives terms cancelling
one another in virtue of Ward identities \cite{barvin}). This deexponentiated
constraint can be extracted from under the integral sign in (\ref{2.13}) in
the form of the differential operator acting on $q_+$,
$\hat{T}_\mu(q_+,\hbar\partial/i\partial q_+)$, so that the equation
(\ref{2.13}) takes the form of the quantum Dirac constraint \cite{barvin0}.
This nonperturbative derivation is, however, purely formal and in an
uncontrolable way depends on the skeletonization of the path integral
\cite{barvin0}. In contrast with these formal considerations in the
following sections we prove that this path integral solves the quantum
constraints at least in the one-loop order of semiclassical expansion.

The Feynman diagrammatic technique for the integral (\ref{2.1}) was also
built in \cite{barvin}. Its main emphasis concerns, certainly, the
boundary conditions at $t_\pm$, because in other respects the
$\hbar$-expansion produces a standard set of Feynman graphs. The first step
in this expansion consists in finding the stationary point of the path
integral subject to boundary conditions (\ref{2.11}) - (\ref{2.12}). As
shown in \cite{barvin}, for gauge fixed action (\ref{2.2}) this is a unique
solution $g=g(\,t\,|\,q_+,q_-)$ of the following boundary value problem
for classical equations of motion in the chosen relativistic gauge\footnote
{
The variation of the gauge fixed action gives for $t_-\leq t\leq t_+$
classical equations amended by the gauge fixing term and two conditions
at the boundaries $t=t_\pm$ (requirement of vanishing coefficients of
$\delta N^\mu(t_\pm)$). In view of (\ref{2.5}) these conditions reduce to
gauge conditions $\chi^\mu=0$ enforced at $t=t_\pm$. On the other hand,
in view of the identity (\ref{1.10}) the classical equations with a gauge
fixing term result in homogeneous equations for these gauge conditions
$Q^\mu_\nu c_{\mu\alpha}\chi^\alpha=0,\,t_-\leq t\leq t_+$. The second order
differential operator $Q^\mu_\nu$ is assumed to be invertible under the
Dirichlet boundary conditions, so that gauge conditions are enforced for
all $t$.
}
	\begin{eqnarray}
	&&\frac{\delta S\,[\,g\,]}
	{\delta g^a(t)}=0,                        \label{2.15}\\
	&&\chi^\mu(g,\dot{g})=0,\,\,\,
	t_-\leq t\leq t_+,                         \label{2.16}\\
	&&q(t_\pm)=q_\pm.                          \label{2.17}
	\end{eqnarray}

Performing the gaussian integration over quantum fields in the vicinity
of this stationary point one arrives at the semiclassical
answer (\ref{0.1}) for our path integral with the tree-level action
	\begin{eqnarray}
	{\mbox{\boldmath$S$}}(q,q')=
	S\,[\,g(\,t\,|\,q_+,q_-)\,]              \label{2.18}
	\end{eqnarray}
and formal expression for the one-loop preexponential factor in terms
of the functional determinants of the ghost and gauge field
operators and the local measure
	\begin{eqnarray}
	{\mbox{\boldmath$P$}}(q_+,q_-)=
	\mu\,\![\,g\,]\! \left.\frac{{\rm Det}\,Q^\mu_\nu}
	{(\,{\rm Det}\, F_{ab}\,)^{1/2}}\,
	\right|_{\,g=g(\,t\,|\,q_+,q_-)}.           \label{2.19}
	\end{eqnarray}
The gauge field operator is given by the second
order variational derivatives of the gauge-fixed action (\ref{2.2}).
In the covariant condensed notations it equals
	\begin{eqnarray}
	&&F_{ab}=S_{ab}-
	\chi^\mu_a\,c_{\mu\nu}\chi^\nu_b,             \label{2.20}\\
	&&S_{ab}\equiv\frac{\,\delta^2 S\,[\,g\,]}
	{\delta g^a\,\delta g^b},                    \label{2.21}\\
	&&\chi^\mu_a\equiv
	\frac{\delta\chi^\mu}{\delta g^a},           \label{2.22}
	\end{eqnarray}
where the functional matrix $\chi^\mu_a$ of linearized gauge conditions
is a first-order differential operator with the delta-function type kernel
	\begin{eqnarray}
	\chi^\mu_a=\chi^\mu_{\;\;a}(d/dt)
	\delta(t-t').                                 \label{2.22a}
	\end{eqnarray}

A formal definition of functional determinants in (\ref{2.19}) is
incomplete unless one specifies a functional space on which they are
calculated. This is equivalent to specifying the boundary conditions
for Green's functions of the gauge and ghost operators $G^{ba}=F^{-1\,ba}$
and $Q^{-1\,\nu}_{\,\mu}$ which participate in the variational equations
for their determinants \cite{Feynman}
	\begin{eqnarray}
	\delta\,{\rm ln}\,{\rm Det}\,F_{ab}=
	G^{ba} \delta F_{ab},\,\,\,\,
	\delta\,{\rm ln}\,{\rm Det}\,Q^\mu_\nu=
	Q^{-1\,\nu}_{\,\mu} \delta Q^\mu_\nu          \label{2.23}
	\end{eqnarray}
and multiloop graphs of semiclassical expansion.

Boundary conditions for the gauge propagator were derived in \cite{barvin}.
They actually follow from the linearized version of the boundary value problem
(\ref{2.15}) - (\ref{2.17}). The propagator $G^{ab}(t,t')$ is
$(n+m)\times(n+m)$ matrix-valued function (cf. eq.(\ref{00.9})). Thus, it
requires $(n+m)$ boundary
conditions at $t_\pm$, $n$ of them obviously being the Dirichlet conditions
for its $a=i$-components ($i=1,...n$), while the rest of them coinciding
with $m$ linearized gauge conditions at $t=t_\pm$
	\begin{eqnarray}
	&&F_{ca}(d/dt) G^{ab}(t,t')=
	\delta^b_c\delta(t-t'),                \label{2.24}\\
	&&G^{ib}(t_\pm,t')=0,                   \label{2.25}\\
	&&\chi^\mu_{\;\;a}(d/dt_\pm)
	G^{ab}(t_\pm,t')=0.                     \label{2.26}
	\end{eqnarray}
The latter belong to Robin type because for relativistic gauges the operator
(\ref{2.22a}) contains derivatives transversal to the boundary. These
boundary conditions have the properties of the BRST-invariance, guarantee
gauge independence of the path integral (\ref{2.1}) \cite{barvin} and
selfadjointness of the gauge field operator recently discussed
in \cite{aveskam}.

Finally, the propagator of ghost fields in (\ref{2.23}) satisfies
in view of (\ref{2.11}) the Dirichlet boundary value problem
	\begin{eqnarray}
	Q^{\,\;\alpha}_\mu(d/dt)\,Q^{-1\,\beta}_{\,\alpha}(t,t')=
	\delta^\beta_\mu,\,\,\,
	Q^{-1\,\beta}_{\,\alpha}(t_\pm,t')=0.     \label{2.27}
	\end{eqnarray}

\section{Local measure, Ward identities and gauge independence}
\hspace{\parindent}
The prescription of boundary conditions for Green's
functions does not, however, uniquely determine the
variation of the determinants (\ref{2.23}). Problem is
that kernels of propagators are not smooth functions of their
arguments, and their irregularity enhances when they are acted upon by
two derivatives contained in $\delta F_{ab}(d/dt)$ and
$\delta Q^{\;\;\mu}_\nu(d/dt)$. Therefore, one has to prescribe the way how
these derivatives act on both arguments of Green's functions and how to take
the coincidence limit of the resulting singular kernels in the functional
traces of (\ref{2.23}). It is well known that strongest divergences
arising from these singular kernels are compensated by the local measure
$\mu\,[\,g\,]$ \cite{Bern}. As far as it concerns the rest finite (or less
divergent in field theories) part, its construction should maintain the
gauge invariance properties encoded in the path integral at nonperturbative
level. This might serve as a hint that can (at least partially) fix the
ambiguities of the above type. For this reason, in this section we consider
the role of local measure and Ward identities in the construction of the
gauge-independent prefactor (\ref{2.19}).

It is well known that strongest power divergences of
functional determinants are related to terms with second order
derivatives in the ghost (\ref{2.7}) and gauge field operators
\cite{Bern,reduct}
	\begin{eqnarray}
	&&F_{ab}(d/dt)=
	-\frac d{dt} a_{ab} \frac d{dt}+...,     \label{2.28}\\
	&&a_{ab}=\frac{\partial^{2}L_{\rm gf}}
	{\partial\dot g^{a}\partial\dot g^{b}}.   \label{2.29}
	\end{eqnarray}
The Hessian matrix of the gauge-fixed Lagrangian is nondegenerate due
to the relativistic nature of gauge conditions (\ref{2.4})-(\ref{2.5}).
The gauge breaking term contributes the part
quadratic in velocities of Lagrange multiplyers $\dot{N}^\mu$ that were
initially absent in the original Lagrangian $L(q,\dot{q},N)$
	\begin{eqnarray}
	L_{\rm gf}(g,\dot g)=L(q,\dot{q},N)-\frac 12
	\chi^\mu(g,\dot g)\,c_{\mu\nu}\chi^\nu(g,\dot g).   \label{2.29a}
	\end{eqnarray}
Therefore, the Hessian matrix has the following components
	\begin{eqnarray}
	&&a_{ik}=G_{ik}-\frac{\partial\chi^\alpha}
	{\partial\dot{q}^i}c_{\alpha\beta}
	\frac{\partial\chi^\beta}
	{\partial\dot{q}^k},                            \label{2.30}\\
	&&a_{i\mu}=\frac{\partial\chi^\alpha}
	{\partial\dot{q}^i}\,c_{\alpha\beta}\,
	a^\beta_\mu,  \label{2.31}\\
	&&a_{\mu\nu}=-a^\alpha_\mu\,c_{\alpha\beta}\,
	a^\beta_\nu,     \label{2.32}
	\end{eqnarray}
where $a^\alpha_\mu$ is the Hessian matrix (\ref{2.5}) of the ghost
Lagrangian. In virtue of the relation
	\begin{eqnarray}
	{\rm det}\,a_{ab}={\rm det}\,G_{ik}\,{\rm det}\,
	c_{\alpha\beta}\,(\,{\rm det}\,a^\mu_\nu\,)^2      \label{2.33}
	\end{eqnarray}
the expression for the local measure (\ref{2.8}) can be rewritten as the
following ratio of the determinants of the gauge field and ghost
Hessian matrices
	\begin{eqnarray}
	\mu\,[\,g\,]=\frac{(\,{\rm Det}\,a_{ab}\,)^{1/2}}
	{{\rm Det}\,a^\mu_\nu}.                          \label{2.34}
	\end{eqnarray}
Therefore the one-loop prefactor (\ref{2.19}) takes the form
	\begin{eqnarray}
	{\mbox{\boldmath$P$}}(q_+,q_-)=
	\left(\frac{{\rm Det}\, F_{ab}}
	{{\rm Det}\,a_{ab}}\right)^{-1/2}
	\left.\frac{{\rm Det}\,Q^\mu_\nu}
	{{\rm Det}\,a^\mu_\nu}\,
	\right|_{\,g=g(\,t\,|\,q_+,q_-)}.              \label{2.35}
	\end{eqnarray}
This form is especially adjusted to the manifest cancellation of strongest
divergent parts of determinants: as shown in \cite{reduct}
the $\delta(0)$-type divergences for a second order differential operator
are cancelled by the functional determinant of the corresponding ultralocal
matrix coefficient of its second order derivatives.

The representation (\ref{2.34}) for the local measure, which is similar to
the ratio of gauge field and ghost functional
determinants in (\ref{2.19}), presents a simplest demonstration of gauge
independence. The numerator and denominator separately depend on the choice
of gauge conditions $\chi^\mu(g,\dot{g})$ (the corresponding matrix
$a^\mu_\nu=-\partial\chi^\mu/\partial\dot{N}^\nu$), but in the ratio
this dependence cancels out and the total local measure (\ref{2.8}) turns
out to be gauge independent\footnote
{
Gauge independent expression for local measure (defined by the Hessian of
the classical gauge action (\ref{2.9})) was first obtained in the context
of Einstein gravity theory \cite{Leutw} and then derived for generic gauge
theory from the canonical path integral \cite{FV}.
}
. The mechanism of gauge independence for the rest part of the one-loop
prefactor is more complicated and is based on Ward identities for gauge and
ghost propagators.

These identities follow from the gauge invariance of the classical action
(\ref{1.10}). The functional differentiation of (\ref{1.10}) shows that on
shell, that is on the background satisfying classical equations of motion, the
functional matrix $S_{ab}$ is degenerate because it has
zero-eigenvalue eigenvectors -- the gauge generators
	\begin{eqnarray}
	R^a_\mu S_{ab}=
	-S_a\frac{\delta R^a_\mu}{\delta g^b}=0 \label{2.35a}
	\end{eqnarray}
(cf. the analogous property (\ref{0.7}) in the canonical context). As a
consequence the gauge operator $F_{ab}$ satisfies the relation
	\begin{eqnarray}
	R^a_\mu F_{ab}=-Q^\alpha_\mu
	c_{\alpha\beta} \chi^\beta_b               \label{2.36}
	\end{eqnarray}
which can be functionally contracted with matrices of the gauge $G^{bc}$
and ghost $Q^{-1\,\mu}_\alpha$ propagators introduced above. Integration
by parts of the derivatives in the operators
$R^a_\mu=R^a_{\;\,\mu}(d/dt)$ and $Q^\alpha_\mu=Q^{\;\,\alpha}_\mu(d/dt)$
does not produce additional surface terms in view of the boundary
conditions for the propagators. Therefore, after
using the equations (\ref{2.24}) and (\ref{2.27}) for gauge and ghost
propagators one arrives at the needed Ward identity
	\begin{eqnarray}
	c_{\alpha\beta}\stackrel{\rightarrow}{\chi}{\!}^\beta_b(d/dt)
	G^{bc}(t,t')=-Q^{-1\,\beta}_{\,\alpha}(t,t')\!
	\stackrel{\leftarrow}
	{R}{\!}^{\;c}_\beta(d/dt').       \label{2.37}
	\end{eqnarray}
Although this identity can be obtained from (\ref{2.36}) by formal
inversion of $F_{ab}$ and $Q^\alpha_\mu$ as finite-dimensional matrices,
we emphasize the necessity of a cautious use of condensed notations with
regard to possible surface terms following from integration by parts. The
result is presented in the form clearly indicating the differential
structure of operators acting on the arguments of Green's function.
In what follows we shall often label the differential operators by arrows
to show the direction (right or left) in which they act.

The obtained Ward identities lead to gauge independence of the one-loop
prefactor (\ref{2.19}) provided that we consistently fix the functional
composition laws for $G^{ba} \delta F_{ab}$ and
$Q^{-1\,\nu}_{\,\mu} \delta Q^\mu_\nu$ in (\ref{2.23}). Let us assume that
the time derivatives of varied differential operators $\delta F_{ab}(d/dt)$
and $\delta Q^\mu_\nu(d/dt)$ are understood as acting in two different ways
	\begin{eqnarray}
	&&\delta\,{\rm ln}\,{\rm Det}\,F_{ab}=
	G^{ba}\!\!\stackrel{\leftrightarrow}
	{\delta F}{\!}_{ab},                            \label{2.39}\\
	&&\delta\,{\rm ln}\,{\rm Det}\,Q^\mu_\nu=
	Q^{-1\,\nu}_{\,\mu}\!\!\stackrel{\leftarrow}
	{\delta Q}{\!}^{\;\;\mu}_\nu.                     \label{2.40}
	\end{eqnarray}
In contrast with the ghost operator, for which both of its derivatives are
acting on {\it one} argument of the Green's function, eq.(\ref{2.39}) here
implies a symmetric action of $\delta F_{ab}(d/dt)$ on both arguments of
$G^{ba}(t,t')$ in the sense that
	\begin{eqnarray}
	L^{(2)}_{\rm gf}=\frac12\varphi^a(t)\!
	\stackrel{\leftrightarrow}
	{F}{\!}_{ab}(d/dt)\,\varphi^b(t)                   \label{2.41}
	\end{eqnarray}
represents the quadratic part of the gauge-fixed Lagrangian in perturbations
of field variables $\varphi^a$. It contains the squares of their velocities
$\dot{\varphi}^a(t)$ rather than their second derivatives (and
$\stackrel{\leftrightarrow}{\delta F}{\!}_{ab}(d/dt)$ obviously implies the
variation of this operator with respect to its background field dependence).

With these conventions the gauge variation of the gauge field and
ghost determinants
	\begin{eqnarray}
	&&\delta_\chi\,{\rm ln}\,{\rm Det}\,F_{ab}=
	\delta_\chi\!\!\stackrel{\leftrightarrow}{F}_{ab}\!G^{ba}\equiv
	-2\,c_{\alpha\beta}\!\stackrel{\rightarrow}{\chi}{\!}^\beta_b\,
	G^{ba}\,
	\delta\!\!\stackrel{\leftarrow}{\chi}{\!}^\alpha_a\nonumber\\
	&&\qquad\qquad\qquad\qquad
	=-2 \int_{t_-}^{t_+} dt\,\left[\,c_{\alpha\beta}
	\!\stackrel{\rightarrow}{\chi}{\!}^\beta_b(d/dt)\,
	G^{ba}(t,t')\,
	\delta\!\!\stackrel{\leftarrow}
	{\chi}{\!}^\alpha_a(d/dt')\,\right]_{\,t=t'},       \label{2.42}\\
	&&\delta_\chi\,{\rm ln}\,{\rm Det}\,Q^\mu_\nu
	=Q^{-1\,\beta}_{\,\alpha}\,\delta_\chi\!\!
	\stackrel{\leftarrow}{Q}{\!}^\alpha_\beta\equiv
	Q^{-1\,\beta}_{\,\alpha}\!
	\stackrel{\leftarrow}
	{R}{\!}^b_\beta\,
	\delta\!\!\stackrel{\leftarrow}{\chi}{\!}^\alpha_b \nonumber\\
	&&\qquad\qquad\qquad\qquad
	=\int_{t_-}^{t_+} dt\,\left[\,
	Q^{-1\,\beta}_{\,\alpha}(t,t')\!
	\stackrel{\leftarrow}
	{R}{\!}^b_\beta(d/dt')\,
	\delta\!\!\stackrel{\leftarrow}
	{\chi}{\!}^\alpha_b(d/dt')\,\right]_{\,t=t'}     \label{2.43}
	\end{eqnarray}
cancel out in virtue of the Ward identities (\ref{2.37}) in the one-loop
prefactor (\ref{2.19}) which proves its gauge independence.
Its independence from the choice of the gauge fixing matrix $c_{\mu\nu}$ is
also based on these identities, though in this case it is cancelled by
the Nielsen-Kallosh factor in the local measure.

For the one-loop effective action these Ward identities were in much detail
analyzed in \cite{PhysRep} in the framework of formal condensed notations.
Here the main emphasis in the mechanism of Ward identities is focused on
boundary conditions and accurate definition of the functional determinants
in (\ref{2.23}). Another choice of the functional composition law in these
variational equations leads in general to extra surface terms breaking the
gauge independence of the one-loop prefactor.

\section{Gauge field functional determinant}
\hspace{\parindent}
Here we calculate the contribution of the gauge field determinant to
the one-loop prefactor (\ref{2.35}). For this purpose we introduce
matrix notations for operators acting in the vector space of
{\it canonical} indices. In these notations the gauge field operator
	\begin{equation}
	{\mbox{\boldmath$F$}}(d/dt)=
	F_{ab}(d/dt)=
	\left[\begin{array}{cc}
        \,F_{ik}(d/dt)\,&\,F_{i\nu}\,(d/dt)\\
        \,F_{\mu k}(d/dt)\, &\, F_{\mu\nu}(d/dt)\,
	\end{array}\right]                          \label{3.1}
	\end{equation}
acts in the space of columns
	\begin{equation}
	\varphi(t)=\left[\begin{array}{c}
        \,\varphi^i(t)\,\\
        \varphi^\mu(t)\end{array}\right].           \label{3.2}
	\end{equation}
Generically it has the form of the second order differential operator
	\begin{equation}
       {\mbox{\boldmath$F$}}\,(d/dt)=-\frac{d}{dt}\,
       {\mbox{\boldmath$a$}}\,\frac{d}{dt}
       -\frac{d}{dt}\,{\mbox{\boldmath$b$}}
	+{\mbox{\boldmath$b$}}^{T}\frac{d}{dt}+
	{\mbox{\boldmath$c$}},                      \label{3.3}
       \end{equation}
where the coefficients
${\mbox{\boldmath$a$}}=a_{ab}(t),\;{\mbox{\boldmath$b$}}=b_{ab}(t)$ and
${\mbox{\boldmath$c$}}=c_{ab}(t)$ are the matrices acting in the vector
space of $\varphi\,(t)$, and the superscript $T$ denotes their
functional transposition $({\mbox{\boldmath$b$}}^{T})\,_{ab}\equiv
b_{ba}$. These coefficients
can be easily expressed as mixed second-order derivatives of the Lagrangian
in the gauge-fixed action (\ref{2.2}) with respect to $g^{a}$
and $\dot{g}^{a}$. In particular, the matrix of second order
derivatives $a\,_{ab}$ is given by the Hessian matrix (\ref{2.29}).

In accordance with the notation (\ref{2.41}) the ``integrated by parts''
version of the operator (\ref{3.1}),
$\stackrel{\leftrightarrow}{{\mbox{\boldmath$F$}}}(d/dt)$, determines the
quadratic part of the gauge fixed Lagrangian with two time derivatives of
${\mbox{\boldmath$F$}}$ acting symmetrically on two functions $\varphi(t)$
and its transpose $\varphi^T(t)$.
For arbitrary two test functions $\varphi_{1}$ and $\varphi_{2}$ the
operator of the form (\ref{3.3}), in our notations, gives rise to the
bilinear form
	\begin{equation}
	\varphi^{T}_{1}\!\stackrel
	{\leftrightarrow}{{\mbox{\boldmath$F$}}}\!\varphi_{2}=
	\dot\varphi^{T}_{1}\,{\mbox{\boldmath$a$}}\,\dot\varphi_{2}+
	\dot\varphi^{T}_{1}\,{\mbox{\boldmath$b$}}\,\varphi_{2}+
	\varphi^{T}_{1}\,{\mbox{\boldmath$b$}}^{T}\dot\varphi_{2}+
	\varphi^{T}_{1}
	{\mbox{\boldmath$c$}}\,\varphi_{2}    \label{3.6}
	\end{equation}
and implies the following integration by parts
           \begin{eqnarray}
	   &&\varphi^{T}_{1}\!\stackrel
	   {\leftrightarrow}{{\mbox{\boldmath$F$}}}\!\varphi_{2}=
	   \varphi^{T}_{1}\,({{\mbox{\boldmath$F$}}}\varphi_{2})+
	   \frac{d}{dt}\left[\,\varphi^{T}_{1}\,
	   ({{\mbox{\boldmath$W$}}}\!\varphi_{2})\,\right],  \label{3.7}\\
	   &&{\mbox{\boldmath$W$}}
	\equiv{\mbox{\boldmath$W$}}(d/dt)
	   ={\mbox{\boldmath$a$}}\,
	\frac{d}{dt}+{\mbox{\boldmath$b$}}.              \label{3.8}
	   \end{eqnarray}
Here ${\mbox{\boldmath$W$}}$ is the {\it Wronskian} operator which enters
the Wronskian relation for ${\mbox{\boldmath$F$}}$
          \begin{equation}
	  \varphi^{T}_{1}\,({\mbox{\boldmath$F$}}\varphi_{2}\!)-
	  ({\mbox{\boldmath$F$}}\varphi_{1}\!)^{T}\varphi_{2}=
	  -\frac{d}{dt}\left[\,\varphi^{T}_{1}\,
	  ({\mbox{\boldmath$W$}}\varphi_{2}\!)-
	({\mbox{\boldmath$W$}}\varphi_{1}\!)^{T}
	  \varphi_{2}\,\right]	                            \label{3.9}
	  \end{equation}
and also participates in the variational equation for the canonical momentum
$\partial L_{\rm gf}/\partial\dot{g}$ valid for arbitrary field variations
$\delta g\,(t)$
	  \begin{eqnarray}
	  \delta\frac{\partial L_{\rm gf}}
          {\partial\dot g}
          ={\mbox{\boldmath$W$}}(d/dt)\,\delta g\,(t).  \label{3.10}
          \end{eqnarray}

Since the velocities of Lagrange multiplyers enter $L_{\rm gf}(g,\dot{g})$
only through the gauge breaking term, the $\mu$-component of the Wronskian
operator is given by
	\begin{eqnarray}
	W_{\mu b}(d/dt)=-\frac{\partial\chi^\alpha}
	{\partial\dot{N}^\mu}c_{\alpha\beta}
	\stackrel{\rightarrow}{\chi}{\!}^\beta_b(d/dt),  \label{3.13}
	\end{eqnarray}
where, in particular, the differential operator of the linearized gauge
conditions (\ref{2.22a}) coincides with that of the boundary value
problem (\ref{2.26}).

Now we introduce the notation for the matrix valued Green's function of
this boundary value problem
	\begin{eqnarray}
	&&G^{ab}(t,t')=
	\left[\begin{array}{cc}
        G^{ik}(t,t')\,&\,G^{i\nu}(t,t')\\
        G^{\mu k}(t,t')\, &\, G^{\mu\nu}(t,t')
	\end{array}\right]
	\equiv{\mbox{\boldmath$G$}}(t,t'),\label{3.15}
	\end{eqnarray}
and in accordance with equations (\ref{2.39}) and (\ref{3.6}) write down
the variation of the gauge field determinant
         \begin{eqnarray}
	 &&\delta\,{\rm ln}\,{\rm Det}\,{\mbox{\boldmath$F$}}
	={\rm Tr}\,\stackrel{\leftrightarrow}
	 {\delta{\mbox{\boldmath$F$}}}\!{\mbox{\boldmath$G$}}=
	\int_{t_{-}}^{t_{+}}dt\,
	 {\rm tr}\left[\,\stackrel{\leftrightarrow}
	 {\delta{\mbox{\boldmath$F$}}}\!
	{\mbox{\boldmath$G$}}\,(t,t^{\prime}\!)\,\right]_
	 {t^{\prime}=t} \nonumber \\
	 \nonumber \\
	 &&\;\;\;\;\;\;\;\;\;\;\;\;\;\;\;\;
	 \equiv\int_{t_{-}}^{t_{+}}dt\,
	 {\rm tr}\left[\,(\delta {\mbox{\boldmath$a$}}\,\frac{d^{2}}{dt\,
	 dt^{\prime}}+\delta {\mbox{\boldmath$b$}}\,\frac{d}{dt^{\prime}}+
	 \delta {\mbox{\boldmath$b$}}^{T}\,\frac{d}{dt}
	+\delta {\mbox{\boldmath$c$}})\,
	 {\mbox{\boldmath$G$}}\,(t,t^{\prime}\!)\,
	 \right]_{t^{\prime}=t}.                       \label{3.16}
	 \end{eqnarray}
Here ${\rm tr}$ denotes the matrix trace operation with respect to condensed
{\it canonical} indices of $\delta {\mbox{\boldmath$a$}}=\delta a_{ab}(t),\,
\delta {\mbox{\boldmath$b$}}=\delta b_{ab}(t),\,\delta
{\mbox{\boldmath$b$}}^{T}=\delta b_{ba}(t),\,
\delta {\mbox{\boldmath$c$}}=\delta c_{ab}(t)$ and
${\mbox{\boldmath$G$}}\,(t,t')={\mbox{\boldmath$G$}}^{ab}\, (t,t')$.

Our goal now will be to construct a special representation
of the Green's function ${\mbox{\boldmath$G$}}(t,t')$ and integrate the
variational equation (\ref{3.16}) by the method of \cite{reduct}.
For this purpose, similarly to \cite{reduct}, we introduce two sets of
basis functions
${\mbox{\boldmath$u$}_{-}}$ and ${\mbox{\boldmath$u$}_{+}}$
of the operator ${\mbox{\boldmath$F$}}$
        \begin{equation}
	{\mbox{\boldmath$F$}}{\mbox{\boldmath$u$}_{\pm}}=0,\,\,
	{\mbox{\boldmath$u_{\pm}$}}=
	{u_{\pm}}^{a}_{A}(t)        \label{3.17}
	\end{equation}
satisfying the full set of boundary conditions (\ref{2.25})-(\ref{2.26})
respectively at $t_{-}$ and $t_{+}$. In view of (\ref{3.13}) and invertibility
of the matrix $\partial\chi^\alpha/\partial\dot{N}^\mu$ the Robin type
boundary conditions (\ref{2.26}) can be rewritten in terms of the
$\mu$-components of the Wronskian operator, so that the full set of
these boundary conditions takes the form
	\begin{eqnarray}
	&&{u_+}^i_A(t_+)=0,\,\,
	\stackrel{\rightarrow}{W}_{\mu a}\!
	{u_+}^a_A(t_+)=0,                             \label{3.18}\\
	&&{u_-}^i_A(t_-)=0,\,\,
	\stackrel{\rightarrow}{W}_{\mu a}\!
	{u_-}^a_A(t_-)=0.                              \label{3.19}
	\end{eqnarray}
In condensed notations we regard these basis functions, enumerated by
the condensed index $A$ of arbitrary
nature, as forming the square matrices with the {\it first} (contravariant)
index $i$ and the {\it second} (covariant) index $A$.

The Wronskian relation (\ref{3.9}) can be used to form the $t$-independent
matrix of the Wronskian inner products
$\Delta_{12}=\varphi^{T}_1({\mbox{\boldmath$W$}}\!\varphi_2)-
({\mbox{\boldmath$W$}}\!\varphi_1)^{T}\varphi_2$ of basis functions
$\varphi_{1,2}={\mbox{\boldmath$u$}}_{\pm}(t)$. In view of the boundary
conditions of the above type this matrix has only two nonvanishing blocks
given by the two mutually transposed matrices
          \begin{eqnarray}
	  &&{\mbox{\boldmath$\Delta$}}_{+-}=
	{\mbox{\boldmath$u^{T}_{+}$}}\,
	  ({\mbox{\boldmath$W$}}{\mbox{\boldmath$u$}}_{-}\!)-
	  ({\mbox{\boldmath$W$}}{\mbox{\boldmath$u$}}_{+}\!)^{T}
	{\mbox{\boldmath$u$}}_{-},\,\,
	{\mbox{\boldmath$\Delta$}}_{+-}
	\equiv({\mbox{\boldmath$\Delta$}}_{+-}\!)_{AB},   \label{3.20}\\
	  &&{\mbox{\boldmath$\Delta$}}_{-+}=
	  {\mbox{\boldmath$u^{T}_{-}$}}\,({\mbox{\boldmath$W$}}
	{\mbox{\boldmath$u$}}_{+}\!)-
	  ({\mbox{\boldmath$W$}}{\mbox{\boldmath$u$}}_{-}\!)^{T}
	{\mbox{\boldmath$u$}}_{+},\,\,
	  {\mbox{\boldmath$\Delta$}}_{-+}
	\equiv({\mbox{\boldmath$\Delta$}}_{-+}\!)_{AB},  \label{3.21}\\
	&&{\mbox{\boldmath$\Delta$}}_{+-}^{T}
	  =-{\mbox{\boldmath$\Delta$}}_{-+}.           \label{3.22}
	  \end{eqnarray}
This fact can also be represented in the form of the following matrix
relation
      \begin{equation}
      \left[\begin{array}{cc}
      \,{\mbox{\boldmath$u$}}^{T}_{-}\,&\,-({\mbox{\boldmath$W$}}\!
	{\mbox{\boldmath$u$}}_{-})^{T}\,\\
      {\mbox{\boldmath$u$}}^{T}_{+}\,&\,-({\mbox{\boldmath$W$}}\!
	{\mbox{\boldmath$u$}}_{+})^{T}
      \end{array}\right]
      \left[\begin{array}{cc}
      \,{\mbox{\boldmath$W$}}\!{\mbox{\boldmath$u$}}_{+}\,&\,
	{\mbox{\boldmath$W$}}\!{\mbox{\boldmath$u$}}_{-}\\
      {\mbox{\boldmath$u$}}_{+}\,&\,{\mbox{\boldmath$u$}}_{-}
	\end{array}\right]=
      \left[\begin{array}{cc}
      \,{\mbox{\boldmath$\Delta$}}_{-+}&0\\
      \,0&{\mbox{\boldmath$\Delta$}}_{+-}\end{array}\right].\label{3.23}
      \end{equation}
Under the assumption of absence of zero modes of ${\mbox{\boldmath$F$}}$
subject to boundary conditions (\ref{2.25})-(\ref{2.26}) (absence of linear
dependence of ${\mbox{\boldmath$u$}}_{+}$ and ${\mbox{\boldmath$u$}}_{-}$)
this relation implies the invertibility of ${\mbox{\boldmath$\Delta$}}_{-+}$.
It also allows one to to establish the following important relations for
equal-time bilinear combinations of basis functions \cite{reduct}
          \begin{eqnarray}
	  &&{\mbox{\boldmath$u$}}_{+}(t)\,
	  ({\mbox{\boldmath$\Delta$}}_{-+}\!)^{-1}
	{\mbox{\boldmath$u$}}_{-}^{T}(t)
	+{\mbox{\boldmath$u$}}_{-}(t)\,
	({\mbox{\boldmath$\Delta$}}_{+-}\!)^{-1}
	{\mbox{\boldmath$u$}}_{+}^{T}(t)=0,         \label{3.24}
	\\
	&&{\mbox{\boldmath$a$}}
	\left[\,\dot{{\mbox{\boldmath$u$}}}_{+}(t)\,
	({\mbox{\boldmath$\Delta$}}_{-+}\!)^{-1}
	{\mbox{\boldmath$u$}}_{-}^{T}(t)
	+\dot{{\mbox{\boldmath$u$}}}_{-}(t)\,
	({\mbox{\boldmath$\Delta$}}_{+-}\!)^{-1}
	{\mbox{\boldmath$u$}}_{+}^{T}(t)\,\right]=
	{\mbox{\boldmath$I$}}.                       \label{3.25}
	\end{eqnarray}
Here ${\mbox{\boldmath$I$}}=\delta^{a}_{b}$ denotes the unity matrix in the
space of canonical indices $a$. To clarify their use in these equations
and in what follows, we
note that in the transposed  matrix ${\mbox{\boldmath$u$}}^{T}_{\pm}=
{\mbox{\boldmath$u_{\pm}$}}_{A}^{a}$ the covariant index $A$ is considered to
be the first
one (in contrast with ${\mbox{\boldmath$u$}}_{\pm}$), so that the matrix
composition law with
$({\mbox{\boldmath$\Delta$}}_{+-}\!)^{-1}=
[\,({\mbox{\boldmath$\Delta$}}_{+-}\!)^{-1}\,]^{AB}$ gives
rise in eqs.(\ref{3.25}) to the matrices with two indices $a$ and
$b$.

By the method of ref.\cite{reduct} (applied there in the case of Dirichlet
boundary conditions) one can show that the Green's function of the mixed
Dirichlet-Robin boundary value problem (\ref{2.24})-(\ref{2.26}) has the
following representation
	\begin{eqnarray}
	&&{\mbox{\boldmath$G$}}\,(t,t')=
	-\theta\,(t\!-\!t')\,
	{\mbox{\boldmath$u$}}_{\!+}(t)\,
	({\mbox{\boldmath$\Delta$}}_{-+}\!)^{-1}
	{\mbox{\boldmath$u$}}_{-}^{T}(t')
	\nonumber\\
	&&\qquad\qquad\qquad\qquad\qquad+\theta\,(t'\!-\!t)\,
	   {\mbox{\boldmath$u$}}_{\!-}(t)\,
	({\mbox{\boldmath$\Delta$}}_{+-}\!)^{-1}
	{\mbox{\boldmath$u$}}_{+}^{T}(t'),       \label{3.26}
	\end{eqnarray}
where $\theta(x)$ is the step function: $\theta(x)=1$ for $x>0$
and $\theta(x)=0$ for $x<0$. This expression is the analogue of
positive-negative frequency decomposition for the Feynman propagator
in scattering theory.

Substituting (\ref{3.26}) to (\ref{3.16}) and repeating the calculations of
\cite{reduct} one can see that in view of relations (\ref{3.24})-(\ref{3.25})
the $\delta(0)$-type terms with derivatives of step functions reduce to the
variation of the logarithm of the local measure for ${\mbox{\boldmath$F$}}$.
Therefore, the variational equation (\ref{3.16}) acquires the form taking
into account the cancellation of $\delta(0)$-type divergences
         \begin{equation}
	 \delta\;{\rm ln}\,\frac{{\rm Det}\,{\mbox{\boldmath$F$}}}
	 {{\rm Det}\,{\mbox{\boldmath$a$}}}=
	 -{\rm tr}\;({\mbox{\boldmath$\Delta$}}_{-+}\!)^{-1}
	\int_{t_{-}}^{t_{+}}dt\,
	{\mbox{\boldmath$u$}}_{-}^{T}
	 \stackrel{\leftrightarrow}{\delta{\mbox{\boldmath$F$}}}
	 {\mbox{\boldmath$u$}}_{+}.                       \label{3.27}
	 \end{equation}
Here we reserve the same notation tr for the trace operation in the space of
indices $A$ enumerating the basis functions of ${\mbox{\boldmath$F$}}$, so
that the matrix multiplication here should read
$[\,({\mbox{\boldmath$\Delta$}}_{-+}\!)^{-1}\,]^{AB}
{\mbox{\boldmath$u$}}_{-B}^{T}\stackrel{\leftrightarrow}
{\delta{\mbox{\boldmath$F$}}}{\mbox{\boldmath$u$}}_{+A}$.

The further transformation of the time integral in (\ref{3.27}) is based on
integration by parts which uses the identities (\ref{3.7}), (\ref{3.9})
and their corollaries obtained by replacing the operators
${\mbox{\boldmath$F$}}$ and ${\mbox{\boldmath$W$}}$ with their variations
$\delta{\mbox{\boldmath$F$}}$ and $\delta{\mbox{\boldmath$W$}}$. This
allows one in a systematic way to reduce the integrand to a total derivative
modulo the terms vanishing in virtue of equations (\ref{3.17}) for basis
functions or their varied versions
$\delta{\mbox{\boldmath$F$}}{\mbox{\boldmath$u$}_{\pm}}=
-{\mbox{\boldmath$F$}}\delta{\mbox{\boldmath$u$}_{\pm}}$. The result boils
down to the contribution of surface terms at $t=t_\pm$. In view of
boundary conditions for ${\mbox{\boldmath$u$}_{\pm}}$ they takes the following
form
	\begin{eqnarray}
	&&\int_{t_{-}}^{t_{+}}dt\,
	{\mbox{\boldmath$u$}}_{-}^{T}
	 \stackrel{\leftrightarrow}{\delta{\mbox{\boldmath$F$}}}
	 {\mbox{\boldmath$u$}}_{+}=
	\left[\,u_-^i\delta({\mbox{\boldmath$Wu_+$}})_i-
	({\mbox{\boldmath$Wu_-$}})_\mu
	\delta u^\mu_+\,\right]_{\,t_+}             \nonumber\\
	&&\qquad\qquad\qquad\quad\qquad\qquad
	+\left[\,({\mbox{\boldmath$Wu_-$}})_i\delta u^i_+
	-u_-^\mu\delta({\mbox{\boldmath$Wu_+$}})_\mu\,
	\right]_{\,t_-},                              \label{3.28}
	\end{eqnarray}
where for brevity we introduced notations
	\begin{eqnarray}
	({\mbox{\boldmath$Wu_\pm$}})_i\equiv\,
	\stackrel{\rightarrow}{W}_{ia}\!u^a_\pm,\,\,\,
	({\mbox{\boldmath$Wu_\pm$}})_\mu\equiv\,
	\stackrel{\rightarrow}{W}_{\mu a}\!u^a_\pm.     \label{3.29}
	\end{eqnarray}
It is useful to rewrite this result in the matrix form
	\begin{eqnarray}
	&&\int_{t_{-}}^{t_{+}}dt\,
	{\mbox{\boldmath$u$}}_{-}^{T}
	 \stackrel{\leftrightarrow}{\delta{\mbox{\boldmath$F$}}}
	 {\mbox{\boldmath$u$}}_{+}=
	\left[\,u^i_-\,\,\,-\!({\mbox{\boldmath$Wu_-$}})_\mu\,\right]
	\left[\begin{array}{c}
	\,\delta({\mbox{\boldmath$Wu_+$}})_i\,\\
	\delta u^\mu_+\end{array}\right]_{\,t_+}   \nonumber\\
	&&\qquad\qquad\qquad\quad\qquad\qquad\qquad\qquad
	-\,\Big[-\!({\mbox{\boldmath$Wu_-$}})_i\,\,\,\,u^\mu_-\,\Big]
	\left[\begin{array}{c}
	\delta u^i_+\\
	\,\delta({\mbox{\boldmath$Wu_+$}})_\mu\,
	\end{array}\right]_{\,t_-}.                     \label{3.30}
	\end{eqnarray}
For brevity here the indices $A,B,...$ enumerating the basis functions are
omitted: they can be regarded as included into subscripts $\pm$ -- the
rule which we shall imply in what follows. Thus all the matrices in this
relation are square (but certainly not symmetric). One index of these
matrices, with respect to which the contraction takes place, is a combination
of $i$ and $\mu$ (being in superscript and subscript positions respectively
or vicy versa), while another index is $A$ encoded as we agreed in $\pm$.

The expression for the time-independent Wronskian matrix
${\mbox{\boldmath$\Delta$}}_{-+}$ can also be rewritten in terms of similar
matrices. When calculated at $t_+$ and $t_-$ it looks respectively as
	\begin{eqnarray}
	&&{\mbox{\boldmath$\Delta$}}_{-+}=
	\left[\,u^i_-\,\,\,-\!({\mbox{\boldmath$Wu_-$}})_\mu\,\right]
	\left[\begin{array}{c}
	\,({\mbox{\boldmath$Wu_+$}})_i\,\\
	u^\mu_+\end{array}\right]_{\,t_+},           \label{3.31}\\
	&&{\mbox{\boldmath$\Delta$}}_{-+}=
	\Big[-\!({\mbox{\boldmath$Wu_-$}})_i\,\,\,\,u^\mu_-\,\Big]
	\left[\begin{array}{c}
	u^i_+\\
	\,({\mbox{\boldmath$Wu_+$}})_\mu\,
	\end{array}\right]_{\,t_-}.                     \label{3.32}
	\end{eqnarray}
The nondegeneracy of ${\mbox{\boldmath$\Delta$}}_{-+}$ guarantees the
invertibility of matrix factors in these two expressions.

Combining the obtained expressions (\ref{3.30}), (\ref{3.31}), (\ref{3.32})
with equation (\ref{3.27}) one can convert its right-hand side into the total
variation
	\begin{eqnarray}
	&&\delta\;{\rm ln}\,\frac{{\rm Det}\,{\mbox{\boldmath$F$}}}
	{{\rm Det}\,{\mbox{\boldmath$a$}}}=
	-\delta\;{\rm tr}\,{\rm ln}\,\left[\begin{array}{c}
	\,({\mbox{\boldmath$Wu_+$}})_i\,\\
	u^\mu_+\end{array}\right]_{\,t_+}\!+
	\delta\,{\rm tr}\,{\rm ln}\,\left[\begin{array}{c}
	u^i_+\\
	\,({\mbox{\boldmath$Wu_+$}})_\mu\,
	\end{array}\right]_{\,t_-}
	=\delta\,{\rm ln}\,{\rm det}\,
	{\mbox{\boldmath$D$}}                       \label{3.32a}
	\end{eqnarray}
of the logarithm of the determinant of the matrix
${\mbox{\boldmath$D$}}={\mbox{\boldmath$D$}}_{ab}$ with indices $a=(i,\mu)$
and $b=(k',\nu)$
	\begin{eqnarray}
	{\mbox{\boldmath$D$}}=\left[\begin{array}{c}
	\,({\mbox{\boldmath$Wu_+$}})_i\,\\
	u^\mu_+\end{array}\right]_{\,t_+}
	({\mbox{\boldmath$\Delta$}}_{-+}\!)^{-1}
	\Big[-\!({\mbox{\boldmath$Wu_-$}})_{k'}
	\,\,\,\,u^\nu_-\,\Big]_{\,t_-}.                     \label{3.33}
	\end{eqnarray}
The block form of this matrix
	\begin{eqnarray}
	&&{\mbox{\boldmath$D$}}=\left[\begin{array}{cc}
	\;{\mbox{\boldmath$S$}}_{ik'}\;\,
	&\,\;X^\nu_i\;\\
	\;\,-X^\mu_{k'}\;\;&\;\;C^{\mu\nu}
	\end{array}\right],                           \nonumber\\
	&&{\mbox{\boldmath$S$}}_{ik'}=
	-({\mbox{\boldmath$Wu_+$}})_i(t_+)
	\,{\mbox{\boldmath$\Delta$}}_{-+}^{-1}\,
	({\mbox{\boldmath$Wu_-$}})_k(t_-)             \label{3.34}\\
	&&X^\nu_i=({\mbox{\boldmath$Wu_+$}})_i(t_+)
	\,{\mbox{\boldmath$\Delta$}}_{-+}^{-1}u^\nu_-(t_-), \label{3.35}\\
	&&X^\mu_{k'}=u^\mu_+(t_+)\,
	{\mbox{\boldmath$\Delta$}}_{-+}^{-1}\,
	({\mbox{\boldmath$Wu_-$}})_k(t_-),         \label{3.36}\\
	&&C^{\mu\nu}=u^\mu_+(t_+)\,
	{\mbox{\boldmath$\Delta$}}_{-+}^{-1}\,
	u^\nu_-(t_-),                                \label{3.37}
	\end{eqnarray}
allows one to rewrite its determinant in the form
	\begin{eqnarray}
	{\rm det}\,{\mbox{\boldmath$D$}}=
	{\rm det}\,\left(\,{\mbox{\boldmath$S$}}_{ik'}
	+X^\mu_i C_{\mu\nu}X^\nu_{k'}\,\right)
	\,{\rm det}\,C^{\mu\nu},\,\,\,C_{\mu\nu}
	=\left(C^{\mu\nu}\right)^{-1}.                     \label{3.38}
	\end{eqnarray}

The notation ${\mbox{\boldmath$S$}}_{ik'}$ chosen for the block (\ref{3.34})
of the matrix ${\mbox{\boldmath$D$}}$ is not accidental. This expression
really yields the matrix of second-order derivatives of the
Hamilton-Jacobi function (\ref{0.6}) with respect to its arguments $q^i$ and
$q^{k'}$. Indeed, this matrix coincides with the derivative of the
canonical momentum $-p_{k'}=\partial{\mbox{\boldmath$S$}}/\partial q^{k'}$
with respect to $q^i$. The momentum is taken at the initial moment
$t_-$ on the classical extremal joining the
points $q$ and $q'$. In view of the relation (\ref{3.10})
	\begin{eqnarray}
	{\mbox{\boldmath$S$}}_{ik'}=
	-W_{k'b'}(d/dt_-)\,
	\frac{\partial g^{b'}(t_-|q_+,q_-)}{\partial q^i_+},  \label{3.39}
	\end{eqnarray}
where the derivative $\partial g^{b}(t|q_+,q_-)/\partial q^i_+$
is given in terms of the Green's function $G^{ab}(t,t')$ (see
ref.\cite{BKif})
	\begin{eqnarray}
	\frac{\partial g^b(t|q_+,q_-)}{\partial q^i_+}=
	-W_{ia}(d/dt_+)\,G^{ab}(t_+,t),                    \label{3.40}
	\end{eqnarray}
whence one gets the equation (\ref{3.34}) after using the basis functions
representation for $G^{ab}(t_+,t)$.

Thus, up to field independent normalization constant, the one-loop
contribution of gauge fields equals
	\begin{eqnarray}
	\left(\,\frac{{\rm Det}\,{\mbox{\boldmath$F$}}}{{\rm Det}\,
	{\mbox{\boldmath$a$}}}\,\right)^{-1/2}=
	{\rm Const}\,\left(\,\frac{{\rm det}\,
	\left(\,{\mbox{\boldmath$S$}}_{ik'}
	+X^\mu_iC_{\mu\nu}X^\nu_{k'}\,\right) }
	{{\rm det}\,C_{\mu\nu}}\,\right)^{1/2}.     \label{3.41}
	\end{eqnarray}
This is a part of expression (\ref{0.12}) for the one-loop prefactor
provided we identify  arbitrary elements
$(X^\mu_i,X^\nu_{k'}, C_{\mu\nu})$ of its {\it canonical} gauge fixing
procedure with the quantities (\ref{3.35}), (\ref{3.36}) and (\ref{3.37})
above. Let us now show that the contribution of ghost fields gives the
remaining part of (\ref{0.12}).

\section{Functional determinant for ghost fields}
\hspace{\parindent}
The ghost operator (\ref{2.6}) is not even formally symmetric because
its right and left action is defined correspondingly on contravariant and
covariant vector fields in the space of gauge indices
	\begin{eqnarray}
	&&Q^\mu_{\;\;\nu}(d/dt)f^\nu(t)=
	\chi^\mu_{\;\;a}(d/dt)\,R^a_{\;\nu}(d/dt)\,f^\nu(t),   \label{4.1}\\
	&&Q^{\;\;\mu}_\nu(d/dt)\,f_\mu(t)=
	R^{\;\;a}_\nu(d/dt)\,\chi^{\;\mu}_a(d/dt)\,f_\mu(t),  \label{4.2}
	\end{eqnarray}
where the form of the first-order differential operators
$R^a_\mu(d/dt)$ and $\chi^\mu_a(d/dt)$ varies depending they are acting
on test functions $(\varphi, f)$ with respect to their condensed
indices $a$ or $\mu$
	\begin{eqnarray}
	&&\chi^\mu_{\;\;a}(d/dt)\,\varphi^a(t)=\left(\frac{\partial\chi^\mu}
	{\partial\dot{g}^a}\frac d{dt}+\frac{\partial\chi^\mu}
	{\partial g^a}\right)\,\varphi^a(t),             \label{4.3} \\
	&&\chi^{\;\;\mu}_a(d/dt)\,f_\mu(t)=
	\left(-\frac{\stackrel{\rightarrow}{d}}{dt}
	\frac{\partial\chi^\mu}
	{\partial\dot{g}^a}+\frac{\partial\chi^\mu}
	{\partial g^a}\right)\,f_\mu(t)                  \label{4.4}
	\end{eqnarray}
and
	\begin{eqnarray}
	&&R^a_{\;\;\mu}(d/dt)\,f^\mu(t)=
	\left(\delta^a_\mu\frac d{dt}+...\right)\,f^\mu(t),  \label{4.5}\\
	&&R^{\;\;a}_\nu(d/dt)\,\varphi_a(t)=
	\left(-\delta^a_\mu
	\frac {d}{dt}+...\right)\,\varphi_a(t).            \label{4.6}
	\end{eqnarray}
Pairs of relations (\ref{4.1})-(\ref{4.2}), (\ref{4.3})-(\ref{4.4}) and
(\ref{4.5})-(\ref{4.6}) obviously differ from one another by integration by
parts in bilinear integral forms with $Q^\mu_\nu$, $\chi^\mu_a$ and
$R^a_\nu$ as functional two-point kernels.

Because of the absence of symmetry the Wronskian relation for $Q^\mu_\nu$
	\begin{eqnarray}
	f_{1 \mu}\,(\stackrel{\rightarrow}{Q}{\!}^\mu_{\;\;\nu} f_2^\nu)-
	(f_{1 \mu}\stackrel{\leftarrow}{Q}{\!}^\mu_{\;\;\nu})\,f_2^\nu
	=-\frac d{dt}\left[ f_{1 \mu}
	(\stackrel{\rightarrow}{W}{\!}^\mu_{\;\;\nu} f_2^\nu)-
	(f_{1 \mu}\stackrel{\leftarrow}
	{W}{\!}^\mu_{\;\;\nu})\,f_2^\nu\right]        \label{4.7}
	\end{eqnarray}
involves two different Wronskian operators
$\stackrel{\rightarrow}{W}{\!}^\mu_{\;\;\nu}$ and
$\stackrel{\leftarrow}{W}{\!}^\mu_{\;\;\nu}$
with the arrows indicating the direction in which the differential
operators are acting. Their actions on test functions
	\begin{eqnarray}
	&&\stackrel{\rightarrow}{W}{\!}^\mu_{\;\;\nu} f^\nu=
	-\frac{\partial\chi^\mu}
	{\partial\dot{g}^a}\,\stackrel{\rightarrow}{R}
	{\!}^a_{\;\;\nu}(d/dt) f^\nu(t)                \label{4.8}\\
	&&f_{\mu}\stackrel{\leftarrow}{W}
	{\!}^\mu_{\;\;\nu}\equiv\,
	\stackrel{\rightarrow}{W}{\!}^{\;\;\mu}_\nu f_{\mu}=
	\left(-\frac{\stackrel{\rightarrow}{d}}{dt}
	\frac{\partial\chi^\mu}
	{\partial\dot{N}^\nu}+\frac{\partial\chi^\mu}
	{\partial N^\nu}\right)\,f_{\mu}(t)=
	\stackrel{\rightarrow}
	{\chi}{\!}^{\;\;\mu}_\nu(d/dt)\,f_{\mu}(t)      \label{4.9}
	\end{eqnarray}
are essentially different from one another and do not differ by a simple
functional transposition of one and the same operator\footnote
{
We shall not introduce for these two Wronskian operators separate notations
and will only distinguish them by the (upper or lower) position of their
indices. The order of indices depends on the direction in which the operator
is acting: when acting to the right its second index (irrespective of
its covariant or contravariant nature) gets contracted with the index of
the test function. For operators acting to the left the order of indices
reverses, so that with these conventions
$\stackrel{\leftarrow}{W}{\!}^\mu_{\;\;\nu}=
\stackrel{\rightarrow}{W}{\!}^{\;\;\mu}_\nu$.
}.
Note, in particular, that the operator (\ref{4.9}) coincides with with
the $\nu$-component of the linearized gauge operator (\ref{4.4}).

With these notations one can again construct the Green's function of
$Q^\mu_\nu$ with Dirichlet boundary conditions in terms of {\it doubled}
set of the right and left basis functions
$r^\mu_\pm(t)$ and $v^\pm_\mu(t)$
	\begin{eqnarray}
	&&\stackrel{\rightarrow}{Q}{\!}^\mu_{\;\;\nu}\,
	r^\nu_\pm(t)=0,\,\,\, r^\nu_\pm(t_\pm)=0,      \label{4.10}\\
	&&\stackrel{\rightarrow}{Q}{\!}_\mu^{\;\;\nu}\,
	v^\pm_\nu(t)=0,\,\,\, v^\nu_\pm(t_\pm)=0.       \label{4.11}
	\end{eqnarray}
The matrix of Wronskian inner products of these basis functions has only the
following nonvanishing block components
	\begin{eqnarray}
	&&\theta^{\;\;+}_-=r^\mu_-\,
	(\stackrel{\rightarrow}{W}{\!}_\mu^{\;\;\nu}v^+_\nu)-
	(r^\mu_-\stackrel{\leftarrow}{W}{\!}_\mu^{\;\;\nu})
	\,v^+_\nu,                                       \label{4.12}\\
	&&\theta^{\;\;-}_+=r^\mu_+\,
	(\stackrel{\rightarrow}{W}{\!}^{\;\;\nu}_\mu v^-_\nu)-
	(r^\mu_+\stackrel{\leftarrow}{W}{\!}^{\;\;\nu}_\mu)
	\,v^-_\nu                                         \label{4.13}
	\end{eqnarray}
participating in the analogue of matrix relations (\ref{3.23})
      \begin{equation}
      \left[\begin{array}{cc}
      \,r^\mu_{-}\,&\,\,\;-(Wr_{-})^\nu\,\\
      \,r^\mu_{+}\,&\,\,\;-(Wr_{+})^\nu\,
      \end{array}\right]
      \left[\begin{array}{cc}
	\,(Wv^{+})_\mu\,\,\,&\,\,(Wv^{-})_\mu\,\\
	v^{+}_\nu\,&\,v^{-}_\nu
	\end{array}\right]=
      \left[\begin{array}{cc}
      \,\theta_-^{\;\;+}\;&\,0\,\\
      \;0&\;\theta_+^{\;\;-}\,\end{array}\right],        \label{4.14}
      \end{equation}
where for brevity we introduced the notations $(Wr_{\pm})^\nu\equiv
r^\mu_\pm\!\stackrel{\leftarrow}{W}{\!\!\!}_\mu^{\;\;\nu}=
\;\stackrel{\rightarrow}{W}{\!\!}^\nu_{\;\;\mu}r^\mu_\pm$ and
$(Wv^{\pm})_\mu=\;\stackrel{\rightarrow}{W}{\!\!}_\mu^{\;\;\nu} v^\pm_\nu$.
Similarly to (\ref{3.24}) these equations imply the equal-time bilinear
relations for ghost basis functions
	\begin{eqnarray}
	&&(Wv^{+})_\mu\,(\theta_-^{\;\;+})^{-1}\,r^\nu_-
	+(Wv^{-})_\mu\,(\theta_+^{\;\;-})^{-1}\,r^\nu_+
	=\delta^\nu_\mu,                       \label{4.15} \\
	&&v^{+}_\mu\,(\theta_-^{\;\;+})^{-1}\,(Wr_-)^\nu
	+v^{-}_\mu\,(\theta_+^{\;\;-})^{-1}\,(Wr_+)^\nu
	=-\delta^\nu_\mu,    \label{4.16}\\
	&&v^{+}_\mu\,(\theta_-^{\;\;+})^{-1}\,r_-^\nu
	+v^{-}_\mu\,(\theta_+^{\;\;-})^{-1}\,r_+^\nu
	=0.                                           \label{4.17}
	\end{eqnarray}

In terms of these basis functions the ghost Green's function equals
	\begin{eqnarray}
	&&Q^{-1\;\nu}_\mu(t,t')=-\theta(t-t')\,v^+_\mu(t)
	\,(\theta_-^{\;\;+})^{-1}\,r_-^\nu(t') \nonumber\\
	&&\qquad\qquad\qquad\qquad\qquad
	+\theta(t'-t)\,v^-_\mu(t)
	\,(\theta_+^{\;\;-})^{-1}\,r_+^\nu(t').    \label{4.18}
	\end{eqnarray}
Substituting this expression to (\ref{2.40}) and again repeating the
calculations of \cite{reduct} one gets the $\delta(0)$-type terms
absorbed by the local measure and arrives at the expression
	\begin{eqnarray}
	&&\delta\;{\rm ln}\,\frac{{\rm Det}\,Q^\mu_\nu}
	{{\rm Det}\,a^\mu_\nu}=
	-\frac12\,{\rm tr}\,(\theta_-^{\;\;+})^{-1}
	\int_{t_-}^{t_+} dt\,r^\nu_-(t)
	\stackrel{\leftarrow}{\delta Q}
	{\!}^{\;\;\mu}_\nu\,v^+_\mu(t)   \nonumber\\
	&&\qquad\qquad\qquad\qquad
	+\frac12\,{\rm tr}\,(\theta_+^{\;\;-})^{-1}
	\int_{t_-}^{t_+} dt\,r^\nu_+(t)
	\stackrel{\leftarrow}{\delta Q}{\!}
	^{\;\;\mu}_\nu\,v^-_\mu(t),               \label{4.19}
	\end{eqnarray}
where the $1/2$ coefficients originate from the symmetric prescription for
the theta function\footnote
{
This prescription is apparently related to symmetric ordering of equal-time
operator products which makes them Hermitian, but we shall not trace back
to the roots of this rule in operator quantization.
}
$\theta(0)=1/2$. It differs from the analogous expression (\ref{3.27}) by
the presence of two terms built of two different sets of right and left
basis functions -- the consequence of asymmetry for the ghost operator as
compared to the symmetric gauge field one. These terms are different here
in contrast with the case of $F_{ab}$ for which they coincide and, thus, lead
to 1/2 coefficients adding up to unity (or, irrespective of the rule
$\theta(0)=1/2$, lead to using the identity $\theta(t-t')+\theta(t'-t)=1$).
Their further transformation repeats the calculations of \cite{reduct} briefly
explained in the previous section and leads to the result
	\begin{eqnarray}
	&&-{\rm tr}\,(\theta_-^{\;\;+})^{-1}
	\int_{t_-}^{t_+} dt\,r^\nu_-(t)
	\stackrel{\leftarrow}{\delta Q}
	{\!}^{\;\;\mu}_\nu\,v^+_\mu(t)         \nonumber\\
	&&\qquad\qquad\qquad\qquad=
	\delta\,{\rm ln}\,{\rm det}\left[\,v^+_\mu(t_-)
	(\theta_-^{\;\;+})^{-1}r^\nu_-(t_+)\,\right]
	+\delta\,{\rm ln}\,{\rm det}\,a^\mu_\nu(t_-),  \label{4.20}\\
	&&{\rm tr}\,(\theta_+^{\;\;-})^{-1}
	\int_{t_-}^{t_+} dt\,r^\nu_+(t)
	\stackrel{\leftarrow}{\delta Q}{\!}
	^{\;\;\mu}_\nu\,v^-_\mu(t)             \nonumber\\
	&&\qquad\qquad\qquad\qquad=
	\delta\,{\rm ln}\,{\rm det}\left[\,v^-_\mu(t_+)
	(\theta_+^{\;\;-})^{-1}r^\nu_+(t_-)\,\right]
	+\delta\,{\rm ln}\,{\rm det}\,a^\mu_\nu(t_+).   \label{4.21}
	\end{eqnarray}
Each of the first terms on the right hand sides of these two relations is
equivalent to the equations (3.19)-(3.20) of the reference \cite{reduct}
for a {\it symmetric} operator subject to Dirichlet boundary conditions.
On the contrary, the terms with ${\rm det}\,a^\mu_\nu(t_\pm)$ are new and
originate from asymmetric action of
$\stackrel{\leftarrow}{\delta Q}{\!}^{\;\;\mu}_\nu$
in the variational definition of the ghost determinant (\ref{2.40}). Using
$\theta_-^{\;\;+}$ and $\theta_+^{\;\;-}$ calculated from
(\ref{4.12})-(\ref{4.13}) respectively at $t_-$ and $t_+$ and taking into
account that $(Wr_\pm)^\mu(t_\pm)=a^\mu_\nu(t_\pm)\dot{r}^\nu(t_\pm)$ one
finds the following expressions for matrices
	\begin{eqnarray}
	&&v^+_\mu(t_-)(\theta_-^{\;\;+})^{-1}r^\nu_-(t_+)
	=-[\,r_-(t_+)\,(\dot{r}_-)^{-1}(t_-)
	\,]^{\,\nu}_{\,\alpha}\;
	[\,a^{-1}(t_-)\,]^{\,\alpha}_{\,\mu},  \label{4.22} \\
	&&v^-_\mu(t_+)(\theta_+^{\;\;-})^{-1}r^\nu_+(t_-)
	=-[\,r_+(t_-)\,(\dot{r}_+)^{-1}(t_+)
	\,]^{\,\nu}_{\,\alpha}\;
	[\,a^{-1}(t_+)\,]^{\,\alpha}_{\,\mu}.   \label{4.23}
	\end{eqnarray}
Here we imply that $r_\pm(t)=r^\mu_\pm(t)$ and
$\dot{r}_\pm(t)=\dot{r}^\mu_\pm(t)$ are square matrices with the
{\it first} index $\mu$ and the {\it second} index encoded in their
subscripts $\pm$. Substituting these expressions to (\ref{4.20})-(\ref{4.21})
and then to (\ref{4.19}) one finds that the variations of
${\rm det}\,a^\mu_\nu(t_\pm)$ cancel out and the result can be functionally
integrated
	\begin{eqnarray}
	&&\frac{{\rm Det}\!\stackrel{\rightarrow}{Q}{\!}^\mu_\nu}
	{{\rm Det}\,a^\mu_\nu}=
	{\rm const}\,\big(\,{\rm det}\,J^\mu_\nu\,
	{\rm det}\,J'^\mu_\nu\,\big)^{-1/2},            \label{4.24}\\
	&&J^\mu_\nu=[\,\dot{r}_-(t_-)\,(r_-)^{-1}(t_+)
	\,]^{\,\nu}_{\,\nu},                           \label{4.25}\\
	&&J'^\mu_\nu=-[\,\dot{r}_+(t_+)\,(r_+)^{-1}(t_-)
	\,]^{\,\nu}_{\,\nu}.                             \label{4.26}
	\end{eqnarray}
up to numerical normalization constant. The coincidence of notations for
the last two matrices with those of Faddeev-Popov operators in
unitary gauges (\ref{0.10})-(\ref{0.11}) is again not accidental. The proof
of these equalities is as follows.

As it follows from the previous section, to compare our one-loop
prexponential factor with that of the solution
(\ref{0.12}) we have to identify the quantities (\ref{3.35})-(\ref{3.36})
with matrices of canonical gauges. The Faddev-Popov matrix corresponding to
the gauge matrix (\ref{3.35}) is therefore
	\begin{eqnarray}
	J^\mu_\nu=X^\mu_i\,\nabla^i_\nu=
	\nabla^i_\nu\,({\mbox{\boldmath$Wu_+$}})_i(t_+)
	\,{\mbox{\boldmath$\Delta$}}_{-+}^{-1}u^\mu_-(t_-). \label{4.27}
	\end{eqnarray}
We show now that this expression equals (\ref{4.25}). For this purpose,
note that the combination $\nabla^i_\nu\,({\mbox{\boldmath$Wu_+$}})_i(t_+)$
can be identified with the linearized constraint under the variation
of canonical coordinates and momenta induced by variations of Lagrangian
variables $\delta g^a(t)=u^a_+(t)$:
	\begin{eqnarray}
	&&\delta q^i(t_+)\equiv u^i_+(t_+)=0,  \label{4.28}\\
	&&\delta p_i(t_+)\equiv W_{ia}^S(d/dt_+)\,u^a_+(t)=
	W_{ia}(d/dt_+)\,u^a_+(t).                       \label{4.29}
	\end{eqnarray}
Here $W_{ia}^S(d/dt)$ is a Wronskian operator built similarly to (\ref{3.10})
but with respect to the original Lagrangian $L(q,\dot{q},N)$ instead of the
gauge-fixed one. The second equality in (\ref{4.29}) follows from the
Robin-boundary conditions on basis functions (\ref{3.18}). Thus, since
$\nabla^i_\mu$ is the momentum derivative of the constraint (\ref{0.4}), the
combination of the above type coincides with the linearized constraint which
in view of (\ref{2.14}) equals the variation of the $\mu$-component of the
classical equations of motion
	\begin{eqnarray}
	\nabla^i_\nu\,({\mbox{\boldmath$Wu_+$}})_i(t_+)=
	\delta T_\mu(q(t_+),p(t_+))=
	-S_{\mu a}(d/dt_+)\,u^a(t_+).                \label{4.30}
	\end{eqnarray}
Then, in virtue of equations (\ref{3.17}) for basis functions and relation
(\ref{4.9}) for the Wronskian operator of $Q^\mu_\nu(d/dt)$
	\begin{eqnarray}
	\nabla^i_\nu\,({\mbox{\boldmath$Wu_+$}})_i(t_+)=
	-\stackrel{\rightarrow}{W}{\!\!}^{\,\;\alpha}_\nu(d/dt_+)\,
	(\,c_{\alpha\beta}\!
	\stackrel{\rightarrow}{\chi}{\!}^\beta_{\,\;a}\,
	u^a_+\,)(t_+).                             \label{4.31}
	\end{eqnarray}
Substitution of this relation to (\ref{4.27}) leads to the expression bilinear
in basis functions $u^a_\pm(t)$ with one of them acted upon by the
operator of linearized gauge conditions. Such an expression can be
transformed by using the Ward identitiy (\ref{2.37}). Substituting the
basis function representations (\ref{3.26}) and (\ref{4.18}) into it at
$t>t'$ one gets
	\begin{eqnarray}
	c_{\alpha\beta}\stackrel{\rightarrow}{\chi}{\!}^\beta_a(d/dt)\,
	u^a_+(t)\,{\mbox{\boldmath$\Delta$}}_{-+}^{-1} u^b_-(t')
	=-v^+_\alpha(t)\,(\theta_-^{\;\;+})^{-1}\,r_-^\beta(t')
	\!\stackrel{\leftarrow}
	{R}{\!}^{\;b}_\beta(d/dt').       \label{4.32}
	\end{eqnarray}
Using this relation with $t=t_+$ and $t'=t_-$ together with (\ref{4.31})
in the right hand side of (\ref{4.27}) and using the Dirichlet boundary
conditions for ghost basis functions one finally arrives at the equation
(\ref{4.25}). Similar proof holds for (\ref{4.26}).

This accomplishes the derivation of the equation (\ref{4.24}). It
represents a reduction
algorithm that expresses the ghost functional determinant in terms of
determinants of the canonical Faddeev-Popov matrices with qualitatively
lower functional dimensionality. Simultaneously this algorithm
relates the gauge fixing procedure in relativistic gauges to the special
unitary gauges of the form (\ref{0.14}) with the matrix (\ref{0.15})
given by (\ref{3.35}). This reduction algorithm together with the
the result for the gauge field determinant (\ref{3.41}) finalizes the
the proof that the path integral with the one-loop prefactor (\ref{2.19})
indeed represents the semiclassical solution (\ref{0.1}), (\ref{0.12}) of
quantum Dirac constraints.

\section{Conclusions}
\hspace{\parindent}
The equality of expressions (\ref{0.12}) and (\ref{2.19}) comprises the
equality of two gauge independent objects -- the function (\ref{0.12})
independent of the unitary gauge conditions $(X^\mu_i,X^\nu_{k,})$ and
the functional (\ref{2.19}) independent of the relativistic ones
$\chi^\mu_a$. The meaning of this seemingly vacuous equation
is that it establishes the mechanism of identical transition from
unitary to relativistic gauge conditions. Such a transition is very
important because it
proves intrinsic unitarity of a manifestly covariant quantization in terms
of Lagrangian path integral. This transition is generally a very nontrivial
procedure, because in contrast with unitary gauges in relativistic
gauges the theory possesses qualitatively different number of propagating
degrees of freedom: gauge modes of $g^a$, Lagrange multiplyers $N^\mu$
as well as ghost fields $(C^\mu,\bar{C}_\nu)$ are dynamical. It is a subtle
mechanism of BRST invariance that guarantees the cancellation of contributions
of all these modes in physical quantities. This mechanism makes the
result equivalent to quantization in the physical sector, that is in
unitary gauge. However, this transition is usually performed at the level
of formal identical transformation in the path integral, which is reached
only in a singular limit $\epsilon\rightarrow 0$ of the so-called
$\epsilon$-procedure of ref. \cite{FV}.

On the contrary, the equivalence of (\ref{0.12}) and (\ref{2.19}) is actually
achieved without any singular limiting operations -- by independently
calculating these quantities in unitary and relativistic gauges and observing
their coincidence when these two sets of gauge conditions are related by
equations (\ref{3.35})-(\ref{3.37}). These relations are nonlocal in time
-- the matrices of unitary gauges and gauge fixing matrix $C_{\mu\nu}$ express
in terms of basis functions of the gauge field operator, nonlocally depending
on its relativistic gauge fixing elements $\chi^\mu_a$ and $c_{\mu\nu}$.

In physical applications, loop expansion of the path integral as a means
of solving quantum Dirac constraints has a very important advantage. This
technique implies that solutions of canonical constraints
(attached to time foliation) admit the {\em spacetime} covariant description
in terms of Feynman diagrams \cite{BKif}. This property is of crucial
importance for correct regularization of inevitable ultraviolet divergences.
This regularization should maintain the covariance in the form not
splitted by time foliation. Spacetime covariance is not manifest due to
the canonical origin of Dirac constraints, but it gets restored
in the proposed calculation technique, because the arising one-loop
functional determinants can be cast into spacetime covariant form by a
suitable choice of relativistic gauge conditions. Important
implication of this technique is the theory of loop effects in
quantum cosmology including, in particular, effective equations for
expectation values in the early inflationary universe \cite{qcr1}.

The aspects of gauge independence considered above are important in
gauge theory applications in spacetimes with boundaries or nontrivial
time foliations. There exists a long list of examples when the calculations
of a formally gauge independent quantity -- one-loop effective action --
gives different results in different gauges \cite{GrifKos,discr}. No
exhaustive explanation for these discrepancies has thus far been given,
and there is a hope that a careful analysis of Ward identities with
regard to boundary conditions can resolve this problem.

The last but not the least problem that belongs to the scope of
our result is the theory of quantum gravitational
tunneling and the physics of wormholes \cite{worm}.
There is a widespread opinion that the predictions of Euclidean quantum
gravity modelling these phenomena have a questionable status due to
the indefiniteness of the Euclidean gravitational action \cite{indef}.
The conclusion drawn in \cite{RubShved} about the mechanism of
transitions with changing spacetime topology is based on the existence
of a negative mode on the wormhole instanton -- a formal
extrapolation of the mechanism which is directly applicable only to
non-gravitational systems \cite{Coleman}. As it was pointed out in
\cite{RubShved} (see also \cite{TanSas}), this issue should be revised
from the viewpoint of
the Wheeler-DeWitt equation. The necessity of such a revision is obvious
because this negative mode belongs to the conformal sector which is not
dynamically independent in the Lorentzian quantum gravity. Its
contribution is, therefore, cancelled by ghost fields in relativistic
gauges. One should expect that similar cancellation should take place
also in Euclidean theory (despite the hyperbolic vs elliptic nature
of field equations in Lorentzian and Euclidean theories, they demand,
after all, the same total number of boundary conditions)\footnote
{
The difference between the Euclidean and Lorentzian settings mainly
concerns the place where the boundary conditions are consistently
imposed: in the hyperbolic case two conditions are fixed at {\it one}
boundary -- initial Cauchy surface, for the elliptic case of Euclidean
theory they should be imposed separately on two different boundaries.
}.
Other difficulties in Euclidean gravity theory related to this issue
also find place in current literature: the lack of strong ellipticity of
the Dirichlet-Robin boundary value problem (\ref{2.24})-(\ref{2.26})
observed in \cite{avesp1} seems to be explained by the
indefiniteness of the Euclidean gravitational action.
The technique proposed here can be regarded as a direct avenue towards the
resolution of these issues -- the problem which is currently under study.

\section*{Acknowledgements}
\hspace{\parindent}
This work was supported by the Russian Foundation for Basic Research
under grants 96-01-00482 and 96-02-16287 and the European
Community Grant INTAS-93-493-ext. Partly this work has been made possible
also due to the support by the Russian Research Project
``Cosmomicrophysics''.

\end{document}